\documentclass[prb]{revtex4} 
\usepackage{amssymb}
\begin{document} 
 
 
\def\a{\alpha} \def\b{\beta} \def\g{\gamma} \def\d{\delta}
\def\e{\varepsilon} \def\th{\theta} \def\k{\kappa} \def\l{\lambda}
\def\m{\mu} \def\n{\nu} \def\p{\pi} \def\c{\chi} \def\s{\sigma}
\def\t{\tau} \def\o{\omega} \def\ph{\phi} \def\G{\Gamma} \def\D{\Delta}
\def\L{\Lambda} \def\Si{\Sigma} \def\O{\Omega}

 
\def\LL{{\cal L}} \def\DD{{\cal D}} \def\GG{{\cal G}} \def\FF{{\cal F}} 
\def\TT{{\cal T}} \def\QQ{{\cal Q}} \def\MM{{\cal M}}  \def\OO{{\cal O}}   
\def\BB{{\cal B}} \def\A{{\cal A}} 

 
\def\bq{{\mathbf q}} \def\bk{{\mathbf k}} 
\def\bs{{\mathbf s}} \def\bx{{\mathbf x}} 
\def\bR{{\mathbf R}} \def\bQ{{\mathbf Q}}
\def\br{{\mathbf r}} \def\bS{{\mathbf S}}  
\def\bqh{\frac{\bq}{2}} \def\brh{\frac{\br}{2}} 
 
 
\def\ra{\rightarrow} 
\def\Ra{\Rightarrow}  
\def\up{\uparrow} 
\def\down{\downarrow} 
\def\Om{\O_m} 
\def\on{\o_n} 
\def\rTO{\sqrt{\frac{T}{\O}}} 
\def\TO{\frac{T}{\O}} 
\def\Vq{V_\bq} 
\def\xik{\xi_\bk} 
\def\pd{\partial}
 
\def\be{\begin{equation}}\def\ee{\end{equation}} 
\def\bea{\begin{eqnarray}}\def\eea{\end{eqnarray}} 
\def\pref#1{(\ref{#1})} 
 
\def\tende#1{\,\vtop{\ialign{##\crcr\rightarrowfill\crcr 
\noalign{\kern-1pt\nointerlineskip} 
\hskip3.pt${\scriptstyle #1}$\hskip3.pt\crcr}}\,} 
 
\def\insertplot#1#2#3#4{\begin{minipage}{#2} 
\vbox {\hbox to #1 {\vbox to #2 {\vfil%
\includegraphics{#4.ps}#3}}} 
\end{minipage}} 
 
\def\ins#1#2#3{\vbox to0pt{\kern-#2 \hbox{\kern#1 #3}\vss}\nointerlineskip}

\title{The low-energy phase-only action in a superconductor: \\
a comparison with the $XY$ model}
 
\author{ L. Benfatto$^{1,2}$, A. Toschi$^1$, and S. Caprara$^1$}

\affiliation{$^1$Dipartimento di Fisica, Universit\`a di Roma 
``La Sapienza'',\\  
and Istituto Nazionale per la Fisica della Materia (INFM) - 
SMC and Unit\`a di Roma 1,\\
Piazzale Aldo Moro, 2 --- I-00185 Roma, Italy.\\
$^2$D\'epartement de Physique , Universit\'e de Fribourg,\\
Chemin du Mus\'ee, 3 ---  CH-1700 Fribourg, Switzerland.}

\date{\today}
  
\begin{abstract}  
The derivation of the effective theory for the phase degrees of
freedom in a superconductor is still, to some extent, an open issue. 
It is commonly assumed that the classical $XY$ model and its quantum
generalizations can be exploited as effective phase-only models. In 
the quantum regime, however, this assumption leads to spurious
results, such as the violation of the Galilean invariance in the
continuum model. Starting from a general microscopic model, in this 
paper we explicitly derive the effective low-energy theory for the
phase, up to fourth-order terms. This expansion allows us to properly 
take into account dynamic effects beyond the Gaussian level, both in
the continuum and in the lattice model. After evaluating the one-loop 
correction to the superfluid density we critically discuss the 
qualitative and quantitative differences between the results obtained 
within the quantum $XY$ model and within the correct low-energy
theory, both in the case of $s$-wave and $d$-wave symmetry of the 
superconducting order parameter. Specifically, we find dynamic
anharmonic vertices, which are absent in the quantum $XY$ model, and 
are crucial to restore Galilean invariance in the continuum model. As 
far as the more realistic lattice model is concerned, in the 
weak-to-intermediate-coupling regime we find that the
phase-fluctuation effects are quantitatively reduced with respect to 
the $XY$ model. On the other hand, in the strong-coupling regime we
show that the correspondence between the microscopically derived
action and the quantum $XY$ model is recovered, except for the 
low-density regime.
\end{abstract}  
  

\maketitle
\section{Introduction}  
\label{intro}
The description of the low-energy dynamics of the phase $\theta$ of
the superconducting order parameter $\Delta=|\Delta|e^{i\theta}$ has 
recently attracted a renewed interest in connection to the
phenomenology of high-$T_c$ superconducting cuprates. These materials 
are characterized by a strongly anisotropic (quasi-two-dimensional) 
crystal structure and by anomalous normal-state properties, which can
be interpreted in terms of preformed Cooper pairs (non-zero amplitude 
$|\Delta|$) without phase coherence (vanishing superfluid density 
$\rho_s$).\cite{review} Although in ordinary (weak-coupling) 
superconductors both $|\Delta|$ and $\rho_s$ vanish at the critical 
temperature $T_c$, due to thermally excited quasiparticles, it is 
possible that $\rho_s$ vanishes when $|\Delta|$ is finite, due to
phase fluctuations. If such is the case, phase fluctuations should
play a role more relevant than expected in ordinary superconductors, 
both near $T_c$ and deeply in the superconducting state. 

A crucial point in the analyses of phase-fluctuation effects is
clearly the choice of the effective model, the most common being 
the classical $XY$ model, as the paradigm for the universality class 
which is relevant to superconductivity.\cite{pf} The $XY$ model, which 
has also been extensively used in contexts such as systems of
resistively shunted Josephson junctions or granular 
superconductors,\cite{chacra} assumes a phase field $\theta_i$ defined 
on a coarse-grained lattice with a lattice constant determined by the 
coherence length $\xi_0$, which sets the length scale above which the 
fluctuations of the amplitude $|\Delta|$ become uncorrelated, and 
$|\Delta|$ may be assumed to be fixed. Thus, at distances larger than 
$\xi_0$, phase fluctuations only are relevant, and can be described in 
terms of a Josephson-like interaction between neighboring sites $<i,j>$,
\begin{equation}  
H_{cl}=\frac{D_0}{4}\sum_{<i,j>}\left(1-\cos\theta_{ij}\right)\simeq  
\frac{D_0}{8}\int d{\bx}\left[(\vec\nabla\theta)^2-\frac{\xi_0^2}{12} 
\sum_{\alpha=x,y}\left({\partial\theta\over\partial\alpha}\right)^4  
+...\right].
\label{eqxy}  
\end{equation}  
Here $\theta_{ij}\equiv \theta_i-\theta_j\simeq |\vec{\nabla}\theta|\xi_0$ 
is the phase difference between nearest-neighboring sites, and $D_0$ 
is the bare coupling constant. Eq. (\ref{eqxy}) is written for a 
two-dimensional system (which is the case relevant for cuprate 
superconductors). In $d=2$, $D_0$ is an energy which measures the 
superfluid stiffness, i.e., the energetic cost to produce phase
variations in the system. In the continuum (Galilean invariant) model 
$D_0=\rho_s(T=0)/m=\rho/m$, where $m$ is the electron mass, $\rho$ is 
the electron density, and we set $\hbar=1$. In the case of a generic 
$d$-dimensional system, we still refer to $D_0$ as to the stiffness, 
even though the corresponding energy scale is $D_0\xi_0^{d-2}$.

Within the $XY$ model, one can determine the depletion of the
superfluid stiffness due to the collective-mode excitations, i.e., to
the anharmonic terms in the gradient expansion of Eq. (\ref{eqxy}). At 
low temperature, the leading correction to the bare stiffness $D_0$
comes from the fourth-order term, and within perturbation theory with 
respect to the harmonic term one gets, e.g. in $d=2$,
\begin{equation}
\label{correz}
D(T)=D_0\left[1-\frac{\xi_0^2}{4}\langle(\vec\nabla\theta)^2\rangle\right],
\end{equation}
where $D_0=D(T=0)$, since $\langle(\vec\nabla\theta)^2\rangle=4T/\xi_0^2D_0$
when evaluated at Gaussian level. This result led to the proposal that 
phase fluctuations, rather than the $d$ wave quasiparticles excitations 
characteristic of the superconductivity in the cuprates, are
responsible for the linear thermal depletion of the superfluid 
density\cite{pf} which is experimentally observed in these materials,
down to very low temperatures (1---5 K).\cite{exp} However, one would 
expect that quantum effects are relevant at such low temperatures,
and, even though the $XY$ model may be reasonably adopted in the 
{\em classical} regime, the investigation of the {\em quantum} regime
is much more involved.

It has been proposed\cite{arun00,noi2} that the quantum effects can
be partially included in the so-called quantum $XY$ model, by deriving
the Gaussian phase-only effective action from a microscopic BCS model,
while obtaining the anharmonic terms through the expansion the 
$\cos\theta_{ij}$ term of Eq. (\ref{eqxy}). As a consequence, the 
interaction terms in the phase, given by powers of 
$\theta_{ij}\simeq \xi_0 |\vec{\nabla}\theta|$, are purely classical, 
whereas the Gaussian propagator is evaluated in the quantum regime. 
However, this approach turns out to be unsatisfactory within many respects. 

Indeed, since in the quantum case $\langle(\vec\nabla\th)^2 \rangle\neq 0$ 
at $T=0$ (see below), Eq. (\ref{correz}) leads to a finite correction 
to $D_0$ even at $T=0$, which, in turn implies $\rho_s(0)\neq\rho$ in 
the continuum model, thus explicitly violating Galilean invariance. 
Therefore, the description of the phase fluctuations by means of the 
quantum $XY$ model misses some important effect within the continuum model. 
A second issue, which is particularly relevant in connection to the 
description of high-$T_c$ superconductors, concerns the limit of strong 
pairing interaction. Indeed, while at weak coupling the amplitude
fluctuations can be safely neglected in deriving the phase-only
theory, as the interaction increases, the fluctuations of the modulus
of the order parameter become intimately connected with the density 
fluctuations.\cite{depalo,noi3} Since the phase and the density are 
conjugate variables, the description of phase fluctuations requires a 
proper treatment of the amplitude fluctuations in the strong-coupling 
regime.

In this paper we present a systematic comparison between the
phase-only model derived microscopically and the quantum and classical
$XY$ models. To this extent, after describing the formal steps to
derive the effective action in various cases (continuum vs lattice
model, neutral vs charged system), we evaluate the one-loop
corrections to $D_0$ coming from the non-Gaussian (anharmonic)
terms. 

We find that quantum effects generate dynamical interaction terms,
absent in the quantum $XY$ model, whose role is crucial, e.g., in
preserving the Galilean invariance in the continuum model, and in
reproducing the expected mapping of the lattice model onto the $XY$
model in the strong-coupling regime. Some of the results on quantum
corrections in the weak-coupling limit where already presented in a
short paper,\cite{noi3} and we provide here a detailed derivation,
since the issue, which involves many subtleties, has been only
partially addressed in the
literature.\cite{arun00,noi2,depalo,noi3,aitchison95,dattu98,aitchison00,kwon,
sharapov,sharapov2,kim} We mainly focus on the quantum ($T=0$) corrections
to the superfluid density, and we comment on the possible extension to
the classical regime. Indeed, besides the restoration of the Galilean
invariance in the continuum system, the microscopic derivation of a
phase-only theory, developed in this paper, allows to demonstrate
clearly that, in general, the $XY$ model leads to an overestimation of
the depletion of the superfluid $\d D$ stiffness, even when the
realistic effects such as the Coulomb screening are taken into account,
with the only exception of the extreme strong-coupling regime where
the $XY$ and the microscopic estimate of $\d D$ tend eventually to
coincide.  For sake of clarity we summarize below our final results,
with reference to the corresponding Sections where they are derived,  
as a guideline for the reader. We
report in tab. I and II (respectively for the continuum and the
lattice models) the estimations for the ratio between the corrections
to the superfluid density, derived within the microscopic models ($\d
D$) and within the $XY$ model ($\d D_{XY}$) in all the cases considered
in this work.

\begin{center}

\begin{table}[htb]
\label{tab1}
\begin{tabular}{||p{3cm}||c|c|c|c||} 

\multicolumn{5}{l}{Continuum models} \\
\hline 
\hline
  & \multicolumn{2}{|c|}{Quantum regime ($T=0$)} & 
\multicolumn{2}{|c||}{Classical regime (high $T$)}  \\
\hline
 & Neutral system & Charged system & Neutral system & Charged system  \\
\hline 

 & & & & \\
\centering{${\d D}/{\d D_{XY}}$} &  $0$  &  0  & 
 ${1}/{(k_F \xi_0)^2}$ & $ {\e_F}/{\e_C} \;\cdot {1}/{(k_F \xi_0)^{d+1}}$ 
 \\
 &(IV B) & (IV B) & (IV C) & (IV C) \\
\hline
\hline
\end{tabular}
\caption{Ratio between the superfluid-stiffness correction  derived within 
continuum microscopic models and the $XY$ model: the classical and
the quantum case are explicitly considered. Here $k_F$ and $\e_F$ are
the Fermi momentum and the Fermi energy respectively, whereas $\e_C$ is a
typical Coulomb energy scale (cfr. Sec. IV C). Note that, quite
generally, $1/(k_F\xi_0) << 1$, i.e. the $XY$ model tends to
overestimate $\d D$).}
\end{table}
\end{center}

\begin{center}

\begin{table}[htb]
\label{tab2}
\begin{tabular}{||p{3cm}||c|c||} 
\multicolumn{3}{l}{Lattice models (neutral and charged system, quantum and
classical regime)} \\
\hline 
\hline
 & Weak Coupling & Strong Coupling   \\
\hline 

 & &  \\
\centering{${\d D}/{\d D_{XY}}$} &  
 $\simeq {1}/{(k_F \xi_0)^2}$ & $ \simeq 1 $  \\
 & (V) & (VI) \\
\hline
\hline
\end{tabular}
\caption{Ratio between the superfluid-stiffness correction derived within 
lattice microscopic models  and the $XY$ model: the classical and the 
quantum case are in this case qualitatively equivalent.}  
\end{table}

\end{center}

The specific issue of the temperature dependence of the superfluid density
in high-$T_c$ superconductors has been analyzed in
Ref. [\onlinecite{arun00,noi2}]. In general, according to Eq.~\pref{eqxy},
one would expect that Coulomb effects, lifting the phase mode to the plasma
frequency (see Sec. II B), completely suppress the contribution
of phase fluctuations to $D(T)$. However, in Ref. [\onlinecite{noi2}] it
has been shown that the $d$-wave symmetry of the order
parameter plays a crucial role in determining the behavior of
the phase mode in the presence of both long-range Coulomb interactions and 
dissipation. Even though here we ignore
dissipative effects, for the sake of completeness we still compare
$s$-wave and $d$-wave superconductors. Once again, the formalism of the
effective action makes this comparison, to some extent, particularly
simple, and of transparent physical interpretation.

The plan of the paper is the following. In Sec. \ref{contine} we
derive the Gaussian phase-only action starting from the continuum model, 
in the absence of long-range Coulomb forces. The case of the charged 
system is dealt with in Sec. \ref{conticha}. Although most of the
results presented in Sec. \ref{conti} are standard, our scope here is
to introduce the formalism and a classification scheme for the
interaction vertices, which turns out to be particularly useful in 
dealing with non-Gaussian terms. In Sec. \ref{latti} we apply the same 
procedure to a lattice model, which does not appear as a trivial 
extension of the continuum model. In Sec. \ref{corco} we calculate the 
anharmonic terms, providing a classification of the diagrams 
(Sec. \ref{cladi}), and determine the one-loop correction to the
stiffness within the continuum model, both at $T=0$
(Sec. \ref{tizero}) and in the classical limit (Sec. \ref{clali}). The 
results are then compared with the results of the $XY$ model. In Sec. 
\ref{corla} the same analysis is carried out in the lattice case. The 
strong-coupling regime of the lattice model is discussed in Sec. 
\ref{stronco}, and the results are summarized in Sec. \ref{concl}. The 
details on the derivation of the phase-only action are reported in 
App. \ref{dedu}, whereas in App. \ref{inva} we discuss the 
connection between the coefficients of the Gaussian action and the 
gauge-invariant electromagnetic response functions.

\section{The effective action for the continuum model}  
\label{conti}
\subsection{The neutral system}
\label{contine}
To introduce the general formalism adopted in this paper, and for the 
sake of completeness, we shortly discuss the standard formal steps to 
derive the phase-only action within the continuum BCS model
$H=H_0+H_I$, with
\bea 
\label{h0} 
H_0&=&-\sum_\sigma\int d\br\frac{1}{2m}
c_\sigma^+(\br)\nabla^2 c_\sigma(\br) ,\\ 
\label{hi}
H_I&=&-\frac{U}{\Omega}\sum_{\mathbf{k},\mathbf{k^{'}},\mathbf{q}} 
w(\bk)w(\bk')
c^{+}_{\mathbf{k}+\mathbf{\frac{q}{2}}\uparrow}
c^{+}_{-\mathbf{k}+\mathbf{\frac{q}{2}}\downarrow}
c_{-\mathbf{k^{'}}+\mathbf{\frac{q}{2}}\downarrow} 
c_{\bk'+\mathbf{\frac{q}{2}}\uparrow}.
\eea 
Most of our results are obtained for generic spatial dimension $d$ of
the system, unless explicitly indicated. In Eqs. \pref{h0}-\pref{hi},
$c^{(+)}_{\sigma}$ is the annihilation (creation) operator of an 
electron of spin $\sigma$, $\Omega$ is the volume, and $U>0$ is the 
pairing interaction strength. Hereafter we assume $\hbar=k_B=1$. The 
factor $w(\bk)$ controls the symmetry of the Cooper-pair wave function, 
e.g., $w(\bk)=1$ for $s$-wave superconductors, and 
$w(\bk)= \cos{2\phi}$, with $\phi = \mbox{atan} ({k_y/k_x})$, for 
$d$-wave superconductors, in the continuum case and for $d=2$. Note
that, in the case of a $d$-wave symmetry, four nodes are present in 
the quasiparticle excitation gap, and the existence of gapless 
excitations is relevant in determining the low-temperature
thermodynamics of the system.
 
We first discuss the case of a neutral system, while the Coulomb 
interaction will be introduced in Sec. \ref{conticha}, to deal with 
the charged case. As customary, the microscopic effective model for 
the collective modes is derived by considering the action
corresponding to the Hamiltonian \pref{h0}-\pref{hi}, within the 
finite-temperature Matsubara formalism, 
\be  
S_{micro}= S_0+S_I= \int^{\beta}_{0} d\tau\left\{\sum_{\mathbf{k}\sigma}
c_{\mathbf{k}\sigma}^{+}(\tau)[\partial_{\tau}+\xi_{\bk}]
c_{\mathbf{k}\sigma}(\tau)d\tau+H_I(\tau)\right\}, \label{smicro} 
\ee 
where $\tau$ is the imaginary time, $\beta=1/T$, and 
$\xi_\bk=\epsilon_\bk-\mu$ is the band dispersion with respect to the 
chemical potential $\mu$, with $\epsilon_\bk=\bk^2/2m$ in the continuum
case. To obtain the effective action in terms of the order-parameter 
collective degrees of freedom, the interaction is decoupled in the 
particle-particle channel by means of the Hubbard-Stratonovich 
transformation, introducing the auxiliary complex field 
$\Delta(x,\tau)$, as explicitly reported in  App. \ref{dedu}.
It is then possible to make the dependence on the phase of $\Delta(\bq,\tau)$ 
explicit in the action, by performing the gauge transformation 
Eq. \pref{fgauge2} on the fermionic fields. Indeed the dependence on 
the phase is eliminated from $S_I$, and made explicit 
in the free part of the action $S_0$, which now reads
\be 
\tilde S_0=S_0+\int dx~\phi^+(x)\hat\Sigma\phi(x), 
\label{s0tilde}
\ee 
with
\be  
{\hat\Sigma}=\left\{\frac{i}{2}\partial_{\tau}\theta(x)+ 
\frac{1}{8m}\left[\vec{\nabla}\theta(x)\right]^2  
\right\}\hat{\tau}_3+\left\{\frac{i}{4m}\vec{\nabla}\theta(x)\cdot   
\stackrel{\leftrightarrow}{\nabla}\right\}\hat{\tau}_0,
\label{sigre}
\ee 
where $\hat\t_i$ are the Pauli matrices and the operator 
$\stackrel{\leftrightarrow}{\nabla}\equiv(\stackrel{\leftarrow}
{\nabla}-\stackrel{\ra}{\nabla})$ acts on the fermionic Nambu spinor 
$\phi(x)$ defined as the column vector $(c_{\up}(x),c_{\down}^{+}(x))$. 
It is worth noting that the field $\theta$ appears in the action only 
through its time  and spatial derivatives, as it is expected for a 
Goldstone (massless) field.  
Moreover we point out that, while in the $s$-wave case the transformation 
\pref{fgauge2} completely eliminates the phase $\th$ from $S_I$, in 
the $d$-wave case a residual dependence survives. Nonetheless, in
App. \ref{dedu}, where we report some more detail, we show that this 
residual dependence can be safely neglected in the {\em hydrodynamic} 
limit, which is the relevant regime when discussing the low-energy 
properties of the collective mode. Thus one can now integrate out the 
fermions, which appear quadratically in $\tilde S_0+S_I$, leading to the 
effective action \pref{sinte}, written in terms of the collective 
variables only. This standard procedure allows one to analyze the 
starting problem \pref{h0}-\pref{hi} from a different point of view. 
Instead of perturbatively expanding the Hamiltonian in powers of the 
coupling $U$, the effective action \pref{sinte} is expanded by
assuming that the fluctuations of the fields around their saddle-point 
values are small, and that the variations of the phase in space and
time are slow. Moreover, if one is interested in the dynamics of the 
phase at low temperature, the fluctuations of the modulus $|\Delta|$
can be neglected. As far as the specific problem of high-$T_c$ 
superconductors is concerned, this approximation seems particularly 
appropriate for underdoped cuprates, which exhibit an anomalously
large binding energy when compared to the critical 
temperature.\cite{review} On the other hand, one should expect that 
modulus fluctuations can still be relevant at distances smaller than 
the typical length scale over which $|\Delta|$ fluctuates. It is then 
natural to introduce a spatial cut-off for phase fluctuations, of the 
order of the coherence length $\xi_0$. At distances greater than
$\xi_0$ the phase fluctuations are the only relevant degrees of
freedom, and a description in terms of the phase-only action becomes 
meaningful. Accordingly, the cut-off $\xi_0$ for the phase-only action
is defined as the characteristic length scale of the spatial decay of
the correlation function for $|\Delta|$. This point is delicate, since 
the naive identification of the coherence length $\xi_0$ with the 
Cooper-pair size $\xi_{pair}$ is meaningless in $d$-wave
superconductor, where $\xi_{pair}$ diverges at $T=0$, due to the
presence of the gapless quasiparticle excitations mentioned above, as
it is discussed in Ref. [\onlinecite{noi4}].
 
Starting from Eq. \pref{sinte} of App. \ref{dedu}, with 
$|\Delta|=|\Delta|(0)+\delta|\Delta|(\bq)$, and neglecting the 
fluctuations of the modulus $\delta|\Delta|(\bq)$, the sum 
$U^{-1}\sum_{q}|\Delta|(q)^2$ reduces to its $q=0$ value. As a 
consequence, the self-energy \pref{eqs} does no longer depend on 
$|\Delta|$. The action \pref{sinte} is then decomposed as 
$S_{neutral} = S_{MF}+S_{eff}(\theta)$, where 
$S_{MF}=|\Delta|^2(0)/U-
\mbox{Tr}\left[\ln{\cal G}_{0}^{-1}\right]$ is the mean-field BCS action,
and $S_{eff}$ is the phase-only
\be
\label{seff}  
S_{eff}(\theta)=\mbox{Tr}\sum_{N=1}^\infty\frac{1}{N}
({\hat\Sigma}\hat{\cal G}_0)^N=\sum_{N=1}^\infty S_{eff}^{(N)}, 
\ee
which takes into account phase fluctuations around the BCS saddle
point. Here $\hat{\cal G}_0$ is 
the BCS matrix defined in Eq. \pref{g00m}, and the   
trace $\mbox{Tr}$ is explicitly given in $k$-space by  
$$
S^{(N)}_{eff}=\frac{1}{N}\sum_{k_1,...,k_N}\mbox{tr}[ 
\hat{{\cal G}_{0}}(k_1)\hat{\Sigma}(k_1-k_2)\times  
\hat{{\cal G}_{0}}(k_2)\hat{\Sigma}(k_2-k_3)\times  
\hat{{\cal G}_{0}}(k_N)\hat{\Sigma}(k_N-k_1)],     
$$
where ${\rm tr}$ is the trace in the Nambu space only and $\hat\Sigma(k)$
 is the Fourier transform of the expression \pref{sigre}. 
Since the self-energy \pref{sigre} contains powers of the 
field  $\theta$ up to the second, the expansion \pref{seff} is not 
directly an expansion in powers of $\theta$ and the terms with the
same power of $\theta$ must be recollected, leading to the expansion
\be  
S_{eff}(\theta)=\sum_{k=2}^\infty S_{k}=\sum_{k=2}^\infty   
\sum_{q_1,\cdots,q_{k-1}}A_k(q_1,\cdots,q_{k-1})\theta(q_1)\cdots 
\theta(q_{k-1})\theta(-q_1-\cdots -q_{k-1}), 
\label{bseff}  
\ee
which is completely equivalent to Eq. \pref{seff}, and includes both the  
Gaussian ($k=2$) and the anharmonic ($k>2$) terms.
  
\begin{figure}[htb]
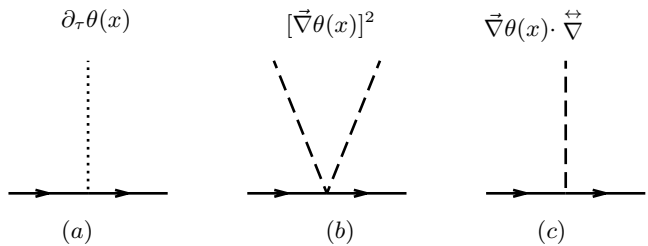
  
\begin{center}  
\insertplot{280pt}{100pt}{
\ins{40pt}{100pt}{$\pd_\t\th(x)$}  
\ins{125pt}{100pt}{$[\vec\nabla\th(x)]^2$}  
\ins{200pt}{100pt}{$\vec{\nabla}\th(x)\cdot \stackrel{\leftrightarrow}  
{\nabla}$}  
\ins{40pt}{20pt}{$(a)$}  
\ins{140pt}{20pt}{$(b)$}  
\ins{220pt}{20pt}{$(c)$}  
}{fig1}
\end{center}  
\caption{Feynman diagrams which correspond to the various insertions of the  
phase field according to Eq. \pref{sigre}. Dotted and dashed lines represent, 
respectively, the insertion of a time derivative or a gradient of the phase
 filed $\th$ in the fermionic lines (depicted with solid lines)}  
\label{fig1}    
\end{figure}

Each term in the effective action \pref{bseff} is given by a closed 
fermionic loop $A_k$ with $k$ incoming lines corresponding to fluctuating 
$\theta$ fields. In forming the fermionic loops, 
according to Eq. \pref{sigre}, we have three kinds of 
fermion-$\theta$ vertices, depicted in Fig. \ref{fig1}:  
\begin{itemize}  
\item[(a)] The vertex with a single incoming dotted line corresponds
to the insertion of the time derivative of the phase, i.e., the term 
$\hat\Sigma_B^1= (i\partial_\tau\theta/2)\hat{\tau}_3$ of Eq. \pref{sigre}.   
\item [(b)] The vertex with two incoming dashed lines corresponds to 
the insertion of the gradient squared of the phase, i.e., the term  
$\hat\Sigma_B^2=[(\vec\nabla\theta)^2/8m]\hat{\tau}_3$ of Eq. \pref{sigre}.  
\item [(c)] The vertex with a single incoming dashed line corresponds to 
the insertion of the spatial derivative of the phase associated to the 
spatial derivative of the incoming and outgoing Green functions, 
i.e., the term  
$\hat\Sigma_F^1=i\{[\vec{\nabla}\theta(x)\cdot
\stackrel{\leftrightarrow}{\nabla}]/4m\}\hat{\t}_0$ of Eq. \pref{sigre}.  
\end{itemize}   

The terms generated in the effective action by the self-energy 
$\hat\Sigma_B= \hat\Sigma_B^1+\hat\Sigma_B^2$, associated to the 
$\hat{\tau}_3$ matrix, are essentially the same which are met when 
deriving the phase-only action in a bosonic superfluid 
system.\cite{popov} We therefore refer to these terms as ``bosonic'' 
contributions. The $\hat \Si_F^1$ term generates instead contributions 
which are strictly related to the existence of quasiparticle
excitations. Therefore, we refer to them as ``fermionic''
contributions. Clearly, the expansion \pref{seff} also generates
``mixed'' contributions, obtained by combinations of $\hat{\t}_0$ and 
$\hat{\t}_3$ terms of the self-energy $\hat \Si$. Such a distinction
will be relevant in the following, while discussing the derivation of 
the anharmonic terms of the phase-only action.
\begin{figure}[ht]  
\begin{center}  
\insertplot{270pt}{150pt}{  
\ins{50pt}{140pt}{$\rho_{MF}/m$}  
\ins{50pt}{20pt}{$\L_{\rho J}$}  
\ins{200pt}{140pt}{$\L_{\rho \rho}$}  
\ins{200pt}{20pt}{$\L_{j j}$}  
}  
{fig2}  
\end{center}  
\caption{Coefficients of the Gaussian action \pref{sgauss}. The first diagram 
comes from the $N=1$ term of Eq. \pref{seff}, while the others come from the 
$N=2$ term of Eq. \pref{seff}. The correspondence between different 
linestyles and the various $\th$ insertions is the same as in Fig. 
\ref{fig1}. Notice that the full current-current correlation function
$\L^{\a\b}_{JJ}=-\rho_{MF}/m\d^{\a\b}+\L^{\a\b}_{jj}$ is
given by the sum of the first and last diagram.}  
\label{figauss}  
\end{figure}  

At Gaussian level $(k=2)$ the effective action describes the collective mode 
of the system. Beyond the Gaussian level $(k> 2)$, interaction terms in the
field $\th$ appear, and we obtain the quantum generalization of the 
anharmonic terms of the $XY$ model, Eq. \pref{eqxy}. As in the classical 
case, the anharmonic terms are responsible for the reduction of the 
stiffness due to phase fluctuations. Before discussing this issue, 
we analyze the structure of the Gaussian phase propagator in the quantum 
regime. The Gaussian effective action is derived with long but 
straightforward calculations, and reads 
\be
S^G_{neutral}= \frac{1}{8}\sum_{q}\left[\Om^2 \L_{\rho\rho}(q)-
\bq_{a}\bq_{b}\L_{JJ}^{ab}(q)+2i\Om\bq_{a}\L_{\rho
J}^{a}(q)\right]|\th(q)|^2. \nonumber \\
\label{sgauss}  
\ee 
The coefficients $\L_{\rho\rho},\L_{\rho J}$ and $ \L_{JJ}$, which are
depicted in Fig.~\ref{figauss} and whose explicit expressions are reported
in App. \ref{dedu}, represent respectively the density-density,
density-current, and current-current bubbles, evaluated with the BCS Green
functions. Since the BCS approximation explicitly breaks the gauge
invariance,\cite{schrieffer} the $\L$ functions do not coincide with the
{\em physical} correlation functions $K$ \cite{sharapov2}, which must obey
the gauge invariance. However, as it is shown explicitly in App. \ref{inva}
(where the general relationship between the $\L$ and the $K$ bubbles is
obtained), it is possible to relate the coefficients of Gaussian effective
action with measurable quantities such as the bare compressibility
$\kappa(T)$ and the stiffness $D(T)=\rho_s(T)/m$ of the system, in the
hydrodynamic limit.\cite{notacompr} Indeed, in this regime, we obtain the
expression 
\be 
S^G_{hydr}=\frac{1}{8}\sum_{q}\left[\kappa_0(T)\Om^2+
D(T)\bq^2\right]\th(q)\th(-q).
\label{sidro}  
\ee 
The Bogoljubov sound mode of the neutral system, with velocity 
$c_{s}=\sqrt{D(T)/\kappa_0(T)}$, is thus recovered. It is worth noting that, 
since the phase and the density are conjugate variables, the collective mode
which appears in the phase-field propagator is the same as the density mode
(see, e.g., Ref. [\onlinecite{anderson}]), as it can be explicitly checked by
deriving the physical density-density correlation function $K_{\rho\rho}$
according to Eq. \pref{eqgaugein}.

By comparing Eq. (\ref{sidro}) with the classical $XY$ model, one can
recognize the same term $\rho_s (\nabla\th)^2$ which appears in
Eq. \pref{eqxy}. However, the quantum Gaussian propagator for phase
fluctuations contains also the $\Om^2$ term of Eq. \pref{sidro} which takes
into account the dynamic effects.  Moreover, according to the results of App. 
\ref{inva} and to Eqs. \pref{eqljj}-\pref{eqcurr}, the temperature 
dependence of the 
bare stiffness $D(T)$ which appears in Eq. \ref{sidro} is
{\em entirely} due to the quasiparticle excitations. Indeed, the coefficient 
$\L_{jj}^{ab}$ in Eqs. \pref{eqljj}-\pref{eqcurr} describes the normal-fluid 
component, whose contribution increases as the temperature increases 
\be
D(T)=\frac{\rho_{MF}}{m}-\frac{2}{\O}\sum_\bk \frac{\bk^2}{m^2}
\left(-\frac{\partial f}{\partial E_\bk}\right)=\frac{\rho_s(T)}{m}.
\label{rhosqp} 
\ee

In the absence of dissipation the coefficients $\L$ are not analytic at $q=0$ 
and finite temperature, since in this case the static 
($\O_m = 0, \bq \ra 0 $) and the dynamic limit ($\bq=0, \O_m \ra 0$) are 
different. This issue has been recently addressed in Ref. 
[\onlinecite{sharapov}]. Since we are interested in the $T=0$
correction to the superfluid density we shall not discuss further the
consequences of such non analyticity, which would disappear in any
case in the presence of dissipation.

\subsection{The charged system}
\label{conticha}  
When dealing with a charged system, the long-range Coulomb interaction 
between the fermions,
\be  
H_{Coul}=\frac{1}{2\O}\sum_{\bk,\bk ',\bq}\sum_{\sigma,\sigma^{'}}V(\bq)   
c^{+}_{\bk+\bq,\s}c^{+}_{\bk '-\bq,\s '}  
c_{\bk ',\s '}c_{\bk,\sigma},  
\label{eqcoul}  
\ee 
must be included in the Hamiltonian \pref{h0}-\pref{hi}. $V(\bq)$ is the 
Fourier transform of the three-dimensional Coulomb repulsion 
$V(r)=e^2/\varepsilon_B r$, projected onto a $d$ dimensional system (e.g., 
$d=2$ for a single layer). Thus, for generic $d$, 
$V(\bq)=\lambda e^2/|\bq|^{d-1}$, where $\lambda$ is a constant which depends 
on the dimension $d$, e.g., $\lambda=4\pi/\varepsilon_B$ for $d=3$ (isotropic 
three-dimensional system) and $\lambda=2\pi/\varepsilon_B$ for $d=2$ (single 
layer). Here $\varepsilon_B$ is the dielectric constant of the ionic 
background. The background ensures overall charge neutrality and 
cancels out the apparent divergence of the Coulomb interaction
$V(\bq)$ at $\bq=0$. We point out that
Eq. \pref{eqcoul}  generically describes a density-density interaction 
mediated by the potential $V(\bq)$. The particular choice of $V(\bq)$ 
distinguishes the short-range from the long-range case.
 
After including the term \pref{eqcoul} in Eq. \pref{smicro}, we 
follow the standard procedure and introduce a Hubbard-Stratonovich
field $\rho_{HS}$ associated to the electron density, 
thus decoupling the Coulomb interaction in the {\em particle-hole} channel. 
Since the short-range pairing in Eq. \pref{hi} is 
important for small center-of-mass momentum, while the Coulomb effects are 
important for small momentum transfer, one can reasonably assume that the 
breakup of the actual interaction in this manner is physically sensible and 
does not lead to any ``overcounting''. Note that in principle also the pairing 
term should be decoupled in the particle-hole channel, but these terms
are safely negligible, at least for weak and 
intermediate $U$. Nevertheless, in  Sec. \ref{stronco}
we shall discuss the subtleties of such a procedure, in the case of 
strong pairing interaction.  

As described in App. \ref{dedu}, by separating 
$\rho_{HS}(q)=\rho_{HS}(0)+\delta\rho_{HS}(q)$, and by neglecting 
the fluctuations of the modulus $\delta|\Delta|$, the expansion of the 
phase-only action around the BCS saddle point leads to 
$S_{charged}  =  S_{MF}+S_{eff}(\th,\rho_{HS})$, where now
\be 
\label{seffc} 
S_{eff}(\th,\d\rho_{HS}) = \sum_q\frac{\d\rho^2_{HS}(\bq)}{2V(\bq)}+ 
\mbox{Tr}\sum_{N=1}^\infty \frac{1}{N}  
({\hat \Si_c}\hat{\cal G}_0)^N, 
\ee 
and
\be 
{\hat \Si_c}= 
\left[\frac{i\partial_{\tau}\th(x)}{2}+\frac{(\vec{\nabla}\th(x))^2}{8m} 
+i\d\rho_{HS}(x) 
\right]\hat{\tau}_3+ 
\left[\frac{i}{4m}\vec{\nabla}\th(x)\cdot  
\stackrel{\leftrightarrow}{\nabla}\right]\hat{\tau}_0.  
\label{sigcre} 
\ee 

The Hubbard-Stratonovich term
$\delta\rho_{HS}^2(0)/V(0)$ does not appear in the mean-field action
$S_{MF}$ as it is supposed to be canceled out, in order
to preserve the charge neutrality of the system, by an equal (and opposite
in sign) contribution coming from the interaction with the ionic
background. This term survives instead in the case of short-range
interaction. Since the self-energy $\hat\Si_c$ depends now also on the
second Hubbard-Stratonovich field $\d\rho_{HS}$, to obtain the phase-only
action, one has to proceed in two steps, as is  explicitly shown in App. 
\ref{dedu}: first, the Gaussian action
in both $\theta$ and $\delta\rho_{HS}$ is derived and, then, 
after integrating out the field $\d\rho_{HS}$, one obtains the 
Gaussian phase-only action 
\be 
S^{G}_{charged}(\th) =  \frac{1}{8}\sum_q \left[\Om^2{\cal L}_{\rho\rho}(q)- 
\bq_{a}\bq_{b}{\cal L}_{JJ}^{ab}(q)+2i\bq_{a}\Om {\cal L}_{\rho J}^{a}(q)  
\right]\th(q)\th(-q), 
\label{sgaussc} 
\ee 
which generalizes Eq. \pref{sgauss} for a charged system. The coefficients 
${\cal L}$, whose definitions are reported in App.\ref{dedu},
are simply the coefficients $\L$ of the unscreened phase-only action 
\pref{sgauss} dressed by the density fluctuations within the 
random-phase  approximation (RPA). This represents the only difference between 
the Gaussian actions  \pref{sgauss} and \pref{sgaussc}, 
which have therefore the same formal
structure. 

The inclusion of Coulomb interaction obviously reflects on the
collective mode of the system. Indeed, the expression of the Gaussian action 
\pref{sgaussc} in the hydrodynamic limit differs from Eq. \pref{sidro} only
in the coefficient of the $\Om^2$ term, since both
$\L_{\rho j}^a$ and ${\cal L}_{\rho j}^a$ vanish in the static limit, 
and ${\cal L}_{JJ}^{ab}$ behaves as $\L_{JJ}^{ab}$, 
whereas ${\cal L}_{\rho\rho}$ gives 
the RPA compressibility of the charged system  
\be 
{\cal L}_{\rho\rho}(\bq,\Om=0)\tende{\bq\ra 0}
\frac{\k_0(T)}{1+V(\bq) \k_0(T)}\tende{\bq\ra 0}\frac{1}{V(\bq)},
\label{eqcompr}
\ee 
which vanishes for $\bq\to 0$. As a consequence, the collective mode 
described by the action 
\be
S^G_{charged} =\frac{1}{8}\sum_{q}\left[\frac{\Om^2}{V(\bq)}+
D(T)\bq^2\right]\th(q)\th(-q),
\label{sidrocd} 
\ee 
is the plasma mode in $d$ dimensions. For example, in the three-dimensional 
case Eq. \pref{sidrocd} reads 
\be  
S^G_{charged}=\frac{1}{8}\sum_{q}\bq^2\left[\frac{1}{4\pi  
e^2}\Om^2+D(T)\right]\th(q)\th(-q) 
\label{sidroc} 
\ee 
and the plasma mode is characterized by a finite energy at zero
momentum, $\Omega_P=\sqrt{4\pi e^2 D(T)}=\sqrt{4\pi e^2 \rho_{s}(T)/m}$. At
$T=0$, where $\rho_s$ coincides with the particle density, $\O_P$ is
exactly the plasma frequency of a three-dimensional charged system.
   
The vanishing of the compressibility 
\pref{eqcompr} in the presence of density-density interactions is peculiar 
of the Coulomb case. If one considers a short-range interaction 
$V(\bq)=\tilde V$, Eq. \pref{eqcompr} simply gives the RPA dressing of  
the bare compressibility
\be 
{\cal L}_{\rho\rho}(\bq,\Om=0)\tende{\bq\ra 0} 
\k(T)=\frac{\k_0(T)}{1+\tilde V \k_0(T)}, 
\label{eqcomprs} 
\ee 
which inserted into Eq. \pref{sgaussc} leads to the Gaussian action 
$$
S^G_{short} =\frac{1}{8}\sum_{q}\left[\frac{\k_0(T)}{1+\tilde V  
\k_0(T)}\Om^2+\frac{\rho_{s}(T)}{m}\bq^2\right]\th(q)\th(-q). 
$$
In the presence of short-range density-density interactions the collective 
mode is still a sound mode (known as Bogoljubov-Anderson
mode\cite{anderson}), with velocity
$c_s'=\sqrt{\rho_s(T)/m\k(T)}=c_s\sqrt{1+\tilde V\k_0(T)}$.  
 
\section{The effective action for the lattice model} 
\label{latti}
We extend the standard procedures discussed in Sec. \ref{conti}, to
deal with the more realistic case of a {\em lattice model}. This extension
does not reduce to a trivial modifications of the continuum results, 
the most important differences being the appearance of vertices with more 
than two incoming dashed lines, and the impossibility to absorb the dressing 
of the fermionic lines in a chemical potential shift.

\begin{figure}[htb]
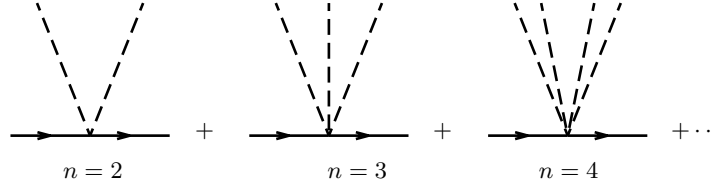
 
\begin{center} 
\insertplot{280pt}{100pt}{ 
\ins{90pt}{35pt}{$+$} 
\ins{180pt}{35pt}{$+$} 
\ins{270pt}{35pt}{$+ \cdots$} 
\ins{40pt}{20pt}{$n=2$} 
\ins{140pt}{20pt}{$n=3$} 
\ins{220pt}{20pt}{$n=4$} 
}{fig3} 
\end{center} 
\caption{Vertices arising from the kinetic term of the lattice model after 
the gauge transformation. Each vertex with $n$ $\th$-lines is associated 
to a same-order derivative of the band dispersion. The first diagram is 
present also in the continuum case (see diagram (b) in Fig. \ref{fig1}), 
with a factor $1/m$. } 
\label{fig3}   
\end{figure}

We first consider the neutral case, the generalization to the
charged case being straightforward. The main difference between the
continuum and the lattice model is in the kinetic term of the
Hamiltonian. Here, we rewrite the free action 
(see Eqs. \pref{h0}, \pref{smicro}) introducing a hopping $t$ between 
nearest-neighboring sites on a lattice of spacing $a=1$,
\be
S_{micro}=\int_0^\b d\t \left\{\sum_{\bx,\s} c_{\bx\s}^+(\t) (\partial_\t
-\m) c_{\bx\s}(\t)-t\sum_{\langle\bx,\bx
'\rangle,\s}[c_{\bx\s}^+(\t) c_{\bx '\s}(\t)+h.c.]+H_I(\t)\right\},
\label{latt} 
\ee 
so that the free-electron dispersion now reads 
$\xi_\bk=-2t\sum_j \cos k_j-\m$, with $j=x_1, \ldots, x_d$. The different 
form of the kinetic term modifies the effect of the gauge transformation on 
the free action $S_0$. As we showed in Sec. \ref{conti}, after the
transformation \pref{fgauge2} the phase disappears from 
$S_I$ and modifies the free action $S_0$ according to
Eq. \pref{s0tilde}. As a consequence, to obtain the self-energy $\hat \Si$
within the lattice model we analyze the effect of the gauge
transformation on the free action \pref{latt}, the
subsequent steps being unchanged. We find two differences with respect to the 
continuum case: 
\begin{itemize} 
\item[(i)] The terms $\hat \Si_F^1$ and $\hat \Si_B^2$ of the self-energy
\pref{sigre} arise, after the gauge transformation, from the kinetic term
$\vec{\nabla}^2/2m$ of the free action. Since the analogous of the velocity
$\bk/m$ of the continuum case is $\partial\xi_\bk/\partial \bk$ in the
lattice case, it follows that for a lattice model each insertion of a
spatial derivative of $\th$ in the fermionic loops is associated to a
same-order $\bk$-derivative of the band dispersion $\xik$.
\item[(ii)] On the lattice the coupling of the fermions to $\th$ induced by
the gauge transformation of the kinetic term generates higher than
second-order derivatives of $\th$ in the self-energy $\hat \Si$. In the
language of Feynman diagrams this means that besides the vertices with one
or two incoming $\th$-lines depicted in Fig. \ref{fig1}, we should add
all the vertices with $n\geq 3$ incoming $\th$-lines, each associated with
a $n$-order $\bk$-derivative of $\xi_\bk$, see Fig. \ref{fig3}.
Observe that no mixing between different components $(a=x,y,\cdots)$ of the 
$\bk$-derivative of $\xi_\bk$ is produced, at least in the case of 
nearest-neighbor hopping.
\end{itemize}

As a consequence of (i) and (ii), the self-energy $\hat\Sigma (k)$
appearing in Eq.~\pref{seff} for $S_{eff}(\theta)$ reads now
\bea 
\hat\Si(k'-k)= 
\rTO \frac{\o_{(k'-k)}}{2}\, \th(k'-k)\hat\tau_3+
\sum_{n=2}^\infty \frac{i^n}{n!}  
\left(\frac{T}{\O}\right)^{n-1} 
\sum_{q_1 \ldots q_n\atop{a=x_1 \ldots x_d}} 
\sin\left(\frac{{\bq_1}_a}{2}\right) \ldots
\sin\left(\frac{{\bq_n}_a}{2}\right)
\times \nonumber\\ \th(q_1) \ldots \th(q_n) \d\left(\sum_{i=1}^n q_i-k'+k\right)
\frac{\partial^n \xi_\bk}{\partial \bk_a^n}\,\hat\tau(n) \label{int} 
\eea
where $\d(q)$ is a delta-function and $\hat\tau(n)$ corresponds to
$\hat\tau_0$ for $n$ odd (fermionic-like contributions) and to $\hat\tau_3$
for $n$ even (bosonic-like contributions).  The functions
$\sin(\bq_i/2)$ in Eq. (\ref{int}) reduce to the factors $\bq_i/2$ of the
continuum case at small momenta, while preserving the lattice periodicity. 
Since we are interested in deriving the hydrodynamic action, in the
following they will be approximated by $\bq_i/2$. 
 
In this section we focus on the generalization of the Gaussian action
derived for the continuum model to the lattice case. To this purpose, it is
sufficient to take into account the modifications described in the item
(i). Indeed, at Gaussian level, only fermionic loops with two $\th$ lines
contribute, i.e., the same diagrams depicted in Fig. \ref{figauss}, which
acquire however a different meaning. In particular, the first fermionic
loop of Fig. \ref{figauss}, which corresponds to the diamagnetic
contribution to the current, does not correspond to the particle density
divided by the mass, resulting instead (in the specific case of a square
lattice with nearest-neighbor hopping) proportional to the average kinetic
energy $E_{kin}$ of the particles
\be 
{\cal T}=\frac{1}{d\O}\sum_{\bk, a} \left(\frac{\partial^2 \xi_\bk}{\partial  
\bk_a^2}\right) \left(1- 
\frac{\xi_\bk}{E_\bk}\right)\tanh \left( \frac{\b E_\bk}{2}\right) = 
a^2 E_{kin} 
\label{d0} 
\ee 
where we indicate with $E_\bk$ the superconducting quasi-particle energy
defined in App. A, and, for the sake of clarity, we made the dependence on
the lattice spacing $a$ explicit. As a consequence, the current-current
correlation function $\L_{JJ}$ becomes now $\L_{JJ}^{ab}= - {\cal T}
\d_{ab}+\L^{ab}_{jj}(q)$, where it is assumed that in the definitions
\pref{eqlrj}-\pref{eqcurr} each factor $\bk_a/m$ is substituted with
$\partial\xik/\partial\bk_a$ and the lattice dispersion $\xik$ appears now
in the quasiparticle energy $E_\bk$.  The stiffness, defined as the static
limit of the transverse part of $\L_{JJ}$, is the lattice analogous of
Eq.~\pref{rhosqp}, i.e.  
\be 
D(T) =\frac{1}{d\O}\sum_{\bk, a}
\left(\frac{\partial^2 \xi_\bk}{\partial \bk_a^2}\right) \left(1-
\frac{\xi_\bk}{E_\bk}\right)\tanh \left( \frac{\b E_\bk}{2}\right)
-\frac{2}{d\O }\sum_{\bk,a} \left(\frac{\partial
\xik}{\partial\bk_a}\right)^2 \left(-\frac{\partial f}{\partial
E_\bk}\right).
\label{stiff} 
\ee 
At Gaussian level the only difference in the phase-only action between the
continuum and the lattice case is in the definitions \pref{rhosqp} or
\pref{stiff} of the superfluid stiffness, the remaining steps being the
same.  In the following, for the sake of simplicity, we write down the
generic Gaussian actions as
\be 
S^G=\frac{1}{8}\sum_{q}\left[\chi\Om^2+ 
D(T)\bq^2\right]\th(q)\th(-q), 
\label{gengauss} 
\ee 
where $\chi$ indicates the static limit of the density-density correlation 
function $\chi(q)\equiv \Lambda_{\rho\rho}(q)$, i.e., the compressibility, 
for neutral  systems
\bea 
\label{chiq}
\chi(q)&=&-\TO \sum_k {\rm tr} 
[\hat{\cal G}_0(k)\hat \t_3\hat{\cal G}_0(k+q)\hat \t_3],\\ 
\label{chi} 
\chi&=& \frac{1}{\O}\sum_\bk \left[ \frac{\D_\bk^2}{E_\bk^3}\tanh \frac{\b 
E_\bk}{2}-2f'(E_\bk)\frac{\xi_\bk^2}{E_\bk^2}\right] \tende{T\ra 0} 
\frac{1}{\O}\sum_\bk \frac{\D_\bk^2}{E_\bk^3}. 
\eea 
which has to be  replaced by $\chi_{LR}(\bq) \sim 1/V(\bq)$ in the  
presence of Coulomb interaction. As a consequence, in the  hydrodynamic 
regime, the Gaussian propagator for phase fluctuations will be generically 
given by
\begin{equation}
P(q)=\langle \th(q)\th(-q)\rangle=4 \left[\chi(q) \Om^2+D\bq^2\right]^{-1}.
\label{pdq}
\end{equation}

The stiffness which appears in the Gaussian propagator is the bare one,
which we indicated with a subscript $D_0$ in Sec. \ref{intro}. However, to
simplify the notation, in the following we drop the subscript, while
keeping in mind that the Gaussian value $D$ is corrected by anharmonic
phase fluctuation according to $D\rightarrow D+\d D$, where $\d D$ will be
calculated within the one-loop approximation.  When quantum corrections are
finite, i.e. $\d D(T=0)\neq 0$, the zero-temperature stiffness does not
coincide with the Gaussian value, defined by the $T=0$ limit of the BCS
formulas \pref{rhosqp} and \pref{stiff}. The consequences of such an
occurrence will be discussed in Sec. \ref{corco}.

\section{One-loop corrections to $D$ within the continuum model} 
\label{corco}
Before deriving the anharmonic terms in the phase-only action, we
discuss the phase-fluctuation correction to the stiffness at $T=0$
in the quantum $XY$ model, which was briefly anticipated in
Sec. \ref{intro}. The quantum $XY$ model includes quantum effects for
the phase through a suitable modification of the Gaussian phase
propagator only, while the interaction terms are powers
of $(\theta_i-\theta_j)\sim \xi_0 |\vec\nabla \theta|$, as in the expansion
(\ref{eqxy}). Consequently
$\d D=-(D\xi_0^2/2d)\langle(\vec\nabla\theta)^2\rangle$, 
which generalizes Eq. (\ref{correz}) in $d$ dimensions. Since 
$\langle(\vec \nabla\theta)^2\rangle=\bq^2 P(q)$, with the quantum propagator 
$P(q)$ defined in Eq. (\ref{pdq}), we find
\bea 
\delta D_{XY}&=&-\frac{D\xi_0^2}{2}\frac{T}{d\Omega} 
\sum_{\bq,\Om}{\mathbf q}^2 P(q) = -2D\xi_0^2\frac{T}{d\Omega}
\sum_{\bq,\Om}\frac{{\mathbf q}^2} {\chi\O_m^2 +D\bq^2}=\nonumber\\ 
&=&-D\xi_0^2 \frac{1}{d\O} \sum_\bq \frac{\bq^2}{\sqrt{\chi D} |\bq|}[1+2b(\b 
  c_s|\bq|)] \tende{T\rightarrow 0}
-D\xi_0^2 \frac{1}{d\O} \sum_\bq \frac{|\bq|}{\sqrt{\chi D}}= 
-\frac{g}{\sqrt{\chi D}}\xi_0^2D, 
\label{xy} 
\eea 
where $b(x)=[e^x-1]^{-1}$ is the Bose function, $c_s=\sqrt{D/\chi}$ is the 
sound velocity, $g={\cal A}_d\zeta^{d+1}/d(d+1)$, ${\cal A}_d$ is the solid angle 
in $d$ dimensions and $\zeta\sim 1/\xi_0$ is the momentum 
cutoff.\cite{notacut} We considered the neutral case as an example, the 
generalization to the charged case being straightforward.

To compare this result with that expected for the microscopically derived
phase-only action, a preliminary distinction arises between {\em continuum} 
and {\em lattice} models. The Gaussian phase propagator is given by
Eq. \pref{gengauss} in both cases, and the temperature dependence of
$D(T)$ is due to quasiparticles only. In spite of this analogy, the 
stiffness $D$ has a different meaning in the continuum or lattice
case. In the former $D=\rho_s/m$, with the constraint  
$\rho=\rho_s+\rho_n$ at any $T$, whereas in the latter $D$ is related to 
the mean kinetic energy, see Eqs. \pref{rhosqp} and \pref{stiff}, 
respectively. In particular, at $T=0$,  
\bea 
\label{cont} 
{\rm continuum~model} &\Ra& D(0)=\frac{\rho_s(0)}{m}=\frac{\rho}{m},\\ 
\label{lattice} 
{\rm lattice~model} &\Ra& D(0)={\cal T}(0)=\frac{1}{d\O}\sum_{\bk, a} \left( 
\frac{\partial^2 \xi_\bk}{\partial \bk_a^2}\right) \left(1- 
\frac{\xi_\bk}{E_\bk}\right).
\eea 
The equality $\rho_s=\rho$
is {\em expected} to hold at $T=0$ in the continuum model, since 
Galilean invariance ensures that at $T=0$, where quasiparticle 
excitations are absent, there is no preferred reference
frame in the system, and all the electrons contribute to the superfluid
component.\cite{abrikosov} Thus, Eq. \pref{cont} states that the 
Gaussian approximations preserves that Galilean invariance of the 
microscopic model. Furthermore, despite the fact that, beyond the
Gaussian level, anharmonic terms induce corrections to the bare
Gaussian stiffness, such corrections are expected to vanish at $T=0$
in the continuum model, to preserve the Galilean-invariant relation
$\rho_s(0)=\rho$. In the lattice case, instead, the presence of finite
corrections to the Gaussian value
$D(0)$ in Eq. \pref{lattice} is not forbidden.  This is exactly what 
happens in the quantum $XY$ model, which is a {\em lattice} model, according to
Eq. (\ref{xy}). As a consequence, the quantum $XY$ model cannot be a
good approximation for the phase-only model of the continuum
system, since it misses the cancellation of the $T=0$ correction to $D$ due to
anharmonic phase fluctuations. This requirement is instead fulfilled 
by expanding the microscopically derived $S_{eff}(\th)$
beyond the Gaussian level, and deriving one-loop corrections to $D$, as we
shall show in the following.

The failure of the quantum $XY$ model for continuum systems is
not a purely formal question, as it rises the issue of its {\em quantitative}
validity both in the lattice case at $T=0$, and in lattice and continuum systems
in the classical regime, above some crossover temperature $T_{cl}$ 
(which depends on the parameter of the microscopic model).  This
issue is of particular relevance when high-$T_c$ superconductors are
considered. Indeed, although in charged systems $T_{cl}$ can be a quite
large energy scale, of the order of the plasma energy, dissipation in
$d$-wave superconductors strongly reduces $T_{cl}$, as discussed in
Ref. [\onlinecite{noi2}], making the classical regime experimentally 
accessible at much lower temperature. This situation then requires a 
critical discussion of the corresponding classical model for phase 
fluctuations, with respect to the standard assumption of considering 
the $XY$ model. Here we do not discuss neither dissipative effects, 
nor the precise definition of $T_{cl}$, but we directly identify 
the classical regime by considering only the $\Om=0$ contribution in 
the Matsubara sums which define the phase-fluctuation corrections $\d D$ 
and $\d D_{XY}$. In the case of the $XY$ model,
by considering Eq. (\ref{xy}), we find that in the classical regime 
\begin{equation}
\delta D_{XY}\approx -2\frac{T}{d\xi_0^{d-2}}, 
\label{xyclass} 
\end{equation}
i.e., $\d D_{XY}(T)\propto T$, as already observed in Sec. \ref{intro}. 
We shall find below that also $\d D$ is linear in $T$, but
with a much smaller coefficient  in the weak-to-intermediate-coupling regime 
(see also Ref. [\onlinecite{kwon}]). 
 
\subsection{Classification of the diagrams}
\label{cladi} 
\begin{figure}[ht]
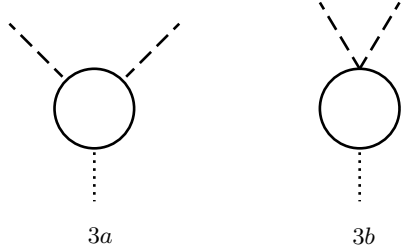
 
\begin{center} 
\insertplot{300pt}{90pt}{ 
\ins{98pt}{5pt}{$3a$} 
\ins{198pt}{5pt}{$3b$} 
} 
{fig4} 
\end{center} 
\caption{Vertices of $S_3$ which contribute to $\d D$. We use the 
same notation of Fig. \ref{fig1}} 
\label{3ver}  
\end{figure}

\begin{figure}[ht]
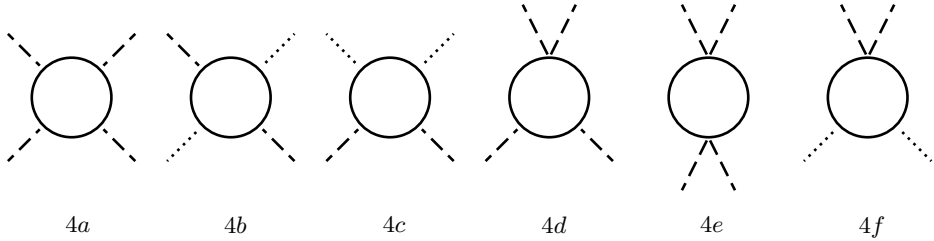
 
\begin{center} 
\insertplot{420pt}{90pt}{ 
\ins{58pt}{5pt}{$4a$} 
\ins{118pt}{5pt}{$4b$} 
\ins{178pt}{5pt}{$4c$} 
\ins{238pt}{5pt}{$4d$} 
\ins{298pt}{5pt}{$4e$} 
\ins{358pt}{5pt}{$4f$} 
} 
{fig5} 
\end{center} 
\caption{Vertices of $S_4$ which contribute to $\d D$. According to our 
classification, $4a$ is a fermionic vertex, $4b,4c,4d$ are mixed 
vertices, and $4e,4f$ are bosonic vertices.} 
\label{4ver} 
\end{figure}

We start our analysis with the continuum  neutral system, while the 
long-range Coulomb forces will be included later. To determine the
one-loop correction $\d D$ to the stiffness  
induced by anharmonic terms in $S_{eff}(\th)$, we derive the  
contributions up to $S_4$ in Eq. (\ref{bseff}). In general, if we separate 
the effective action in the Gaussian ($S^G$) and anharmonic ($S_{anh}$) part,
and make a Taylor expansion of the latter, by means of the Wick theorem 
we can evaluate the two-field propagator as the 
resummation of all the connected diagrams obtained by fixing two $\th$  
legs and contracting the others with the Gaussian propagator $P(q)$,
\bea 
\langle\th(q)\th(-q)\rangle & = & \frac{\int \th(q)\th(-q) e^{-S^G(\th)} 
e^{-S_{anh}(\th)} {\cal D} \th}{Z} \nonumber\\ 
& = & \sum_{n}\frac{(-1)^n}{n!}\frac{\int \th^2 e^{-S^G(\th)} 
[S_{anh}(\th)]^n {\cal D} \th}{Z} 
=  \sum_{n} \frac{(-1)^n}{n!} \langle\th(q)\th(-q) [S_{anh}(\th)]^n 
\rangle_{CON}, \nonumber
\eea 
where $\langle \cdots \rangle_{CON}$ indicates connected diagrams only.  
All one-loop corrections to $P(q)$, i.e., diagrams 
with only one integration on bosonic momenta, are obtained by approximating  
$\sum_n (-1)^n/n![S_{anh}]^n \simeq S_4+[S_3]^2/2$.  
Moreover, since we are specifically interested in the correction to
$D$, which is the coefficient of the gradient term in the effective 
Gaussian action, we select the diagrams which have at least two 
$\vec\nabla \th$ lines, as depicted in Figs. \ref{3ver} and \ref{4ver}. 
 
\begin{figure}
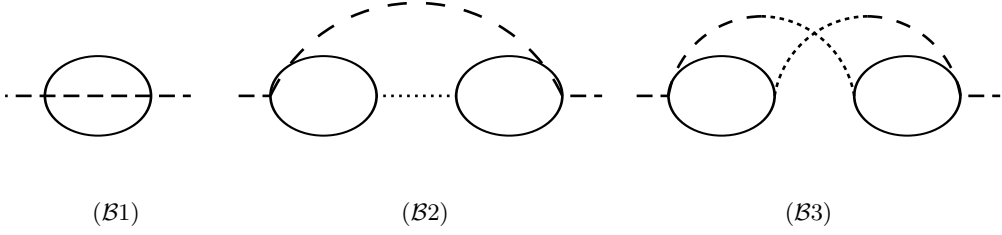
 
\begin{center} 
\insertplot{400pt}{120pt}{ 
\ins{38pt}{40pt}{(${\cal B}1$)} 
\ins{155pt}{40pt}{(${\cal B}2$)} 
\ins{300pt}{40pt}{(${\cal B}3$)} 
} 
{fig7} 
\end{center} 
\caption{Bosonic corrections to $D$. ${\cal B} 1$ arises from the diagram $4e$ 
of Fig. \ref{4ver}, ${\cal B}2$ and ${\cal B}3$ from two $3b$ vertices of 
Fig. \ref{3ver}. Notice that the dashed-dotted line of ${\cal B}3$ indicates 
the average  $\langle\Om\bq \th(q)\th(-q)\rangle$.} 
\label{bos} 
\end{figure} 
 
To obtain one-loop corrections to $D$ we consider all the
possible contractions of two $\th$ lines in one of the fourth-order vertices 
in Fig. \ref{4ver}, or in the combination of two third-order vertices 
in Fig. \ref{3ver}. The above classification in bosonic, fermionic and  
mixed contributions turns out to be particularly convenient while evaluating  
all the resulting one-loop corrections to $D$. 
We can indeed distinguish: 
\begin{itemize} 
\item Fermionic corrections, which are given by the self-energy- and 
vertex-like diagrams obtained from the $4a$ loop in Fig. \ref{4ver}.
\item Bosonic corrections, which arise from two loops of the $3b$ 
type and from the loop $4e$ of Fig. \ref{4ver}, see Fig. \ref{bos}.  
\item Mixed corrections, which are the remaining combinations of 
third- and fourth-order diagrams.
\end{itemize} 

The distinction between fermionic, bosonic, and mixed contributions to $\d
D$ is motivated a posteriori by the fact that {\em all} the contributions
to $\d D$ coming from the fermionic and mixed diagrams cancel out each
other at $T=0$, whatever is the symmetry of the gap and both in continuum
and lattice models. This result is connected to the fact that the
temperature dependence of the fermionic and mixed bubbles is controlled by
the thermal excitation of the quasiparticles, which are suppressed at low
temperature. However, while for $s$-wave superconductors this suppression is
exponential in temperature due to the presence of a finite gap in the
excitation spectrum, for a $d$-wave superconductor the presence of nodal
quasiparticles obliges a careful evaluation of the fermionic contribution
to $\d D$. For example, one can see that the fermionic contributions
generated by the vertex (4a) in Fig. \ref{4ver} can be written as
\be
\d D_\QQ =
\frac{1}{8d}\frac{T}{\O}\sum_{\bq,\Om} P(q) \bq^2\QQ(\bq,\Om),
\label{ddq}
\ee
where $v_\bk=\partial \xi_\bk/\partial \bk$ and 
the hydrodynamic limit of the $\QQ$ bubble is
\be {\cal Q}(\bq\ra
0,\Om=0)=\frac{1}{\O}\sum_\bk v_\bk^4 f'''(E_\bk),
\label{ff3}
\ee 
where $f'''(x)$ is the third-order derivative of the Fermi function. In
the $s$-wave case the integral in Eq. \pref{ff3} vanishes exponentially 
at $T=0$, while in the
$d$-wave case, by evaluating the contribution of nodal quasiparticles to
Eq.  \pref{ff3}, which is the relevant contribution at low temperature, one
finds that it diverges as $1/T$. This can be easily seen by considering
that the density of states of nodal quasiparticles is linear in energy
$N(E)\propto E$, \cite{qp} so that ${\cal Q}(q=0)\propto \int dE N(E)
f'''(E) \ra 1/T$ as $T\ra 0$.  It should be stressed that the divergence in
the $d$-wave case of {\em all} fermionic loops with $N\ge 4$ Green
functions, when evaluated at zero external momenta, is a characteristic
feature of systems having Dirac quasiparticles. Nevertheless the one-loop
expansion of $D$ is well defined: indeed by retaining all the $\bq,\Om$
dependence on the fermionic bubble ${\cal Q}$ one has no longer to deal
with any divergence of fermionic corrections to $D$, which 
result even vanishing in the zero-temperature limit. \cite{samokhin} 

\begin{figure}[ht]
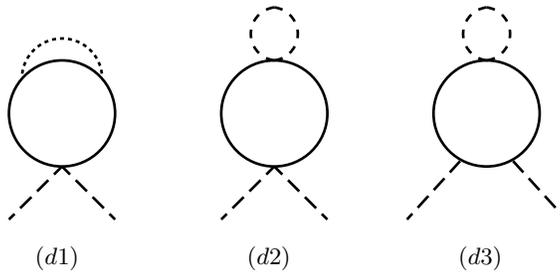
 
\begin{center} 
\insertplot{250pt}{120pt}{ 
\ins{30pt}{20pt}{($d1$)} 
\ins{110pt}{20pt}{($d2$)} 
\ins{190pt}{20pt}{($d3$)} 
} 
{fig9} 
\end{center} 
\caption{Bosonic ($d1$ and $d2$) and mixed ($d3$) one-loop corrections to  
$D$ arising from phase-fluctuation correction of the electron  
density beyond Gaussian level. Working at fixed density, these terms are  
canceled out by the $\d D_\mu$ contribution due to the chemical-potential 
shift.} 
\label{chem} 
\end{figure}

Special attention must be devoted also to the bosonic ($d1$ and
$d2$) and fermionic ($d3$) diagrams depicted in Fig. \ref{chem}. Indeed, 
these diagrams appear as corrections to the bare Green function in
the diagrams of Fig. \ref{figauss} 
which define the the mean-field current-current correlation
function $\L_{JJ}$ in Eq.  \pref{eqljj}. In particular, $d1$ and $d2$ in
Fig. \ref{chem} are phase-fluctuation corrections to the diamagnetic term
$-\rho_{MF}/m$, and $d3$ in Fig. \ref{chem} is a phase-fluctuation
correction to the bubble $\L_{jj}$, which gives the quasiparticle depletion
of $\rho_s(T)$ at finite temperature.  One can easily see that these terms
are due to the shift of $\rho$ with respect to the mean-field value
$\rho_{MF}$, as induced by anharmonic phase fluctuations. On the other
hand, working at fixed density, one assumes $\rho_{MF}\equiv \rho$ and
takes into account these contributions by shifting the chemical potential
$\mu\ra \mu+\d\mu$ with respect to the mean-field value, so that
\be
\d \rho=\rho-\rho_{MF}= 
-\frac{\partial \Delta F}{\partial \mu}+\frac{\partial \rho}{\partial \mu} \d \mu=0 ~~~
\label{mush2} 
\Ra ~~~\d\mu=\frac{1}{\chi}\frac{\partial \Delta F}{\partial \mu}, 
\ee 
where we use the definition
$\chi=\partial \rho/\partial \mu$ for the compressibility, and the free-energy 
contribution $\Delta F$ due to phase fluctuations is evaluated 
within the Gaussian approximation, 
\be 
\Delta F=-\ln \int {\cal D}\th e^{-S^G(\th)}=\frac{1}{2}\TO\sum_q 
\ln [\chi \Om^2+ D\bq^2]. 
\label{freen} 
\ee 
While evaluating one-loop corrections to $D$, one 
takes into account the chemical potential shift \pref{mush2} by adding the 
contribution
\be 
\d D_\mu=\frac{\partial D}{\partial \mu}\d \mu, 
\label{ddmu} 
\ee 
which cancels out exactly the contribution of the diagrams in
Fig. \pref{chem}. This can be easily understood, e.g., at
$T=0$. Because  at $T=0$ 
all the mixed diagrams vanish, only $d1$ and $d2$ contribute 
to $\d D$. Let us consider the diagram $d2$. Its contribution to $\d D$ is
$$
\d D_{d2}=\frac{1}{8m^2}\TO \sum_q \bq^2 P(q)\left\{ \TO\sum_k{\rm tr}  
[\hat{\cal G}_0(k)\hat \t_3\hat{\cal G}_0(k)\hat \t_3]\right\}=-\frac{\chi}{8m^2}\TO 
\sum_q \bq^2 P(q).  
$$
Indeed, the fermionic loop of $d2$, given by the trace in the previous 
equation,  corresponds to the hydrodynamic limit of the same
density-density bubble $\chi(q)$ which appears as the coefficient of the 
$\Om^2$ term in the Gaussian propagator $P(q)$ (\ref{pdq}), see
Eqs.\pref{chiq}-\pref{chi}.\cite{notamat}

On the other hand, while evaluating $\partial \Delta F/\partial \mu$
according to Eq. \pref{freen}, one finds that the $D\bq^2$ term of
Eq. \pref{freen} contributes with
$$ 
\frac{\partial \Delta F}{\partial \mu}= \cdots + \frac{1}{2}\TO \sum_q
\frac{1} {\chi \Om^2+D\bq^2} \bq^2 \frac{\partial D}{\partial \mu}
=\cdots+\frac{\chi}{8m}\TO \sum_q \bq^2 P(q),
$$
where we used Eq. (\ref{pdq}) and $D=\rho_s/m=[\rho-\rho_n(T)]/m$ so that,
at $T=0$, $D=\rho/m$ and $\partial D/\partial\mu=\chi/m$. As a consequence,
by means of Eqs. \pref{mush2} and \pref{ddmu} we find that
$$
\d D_\mu=\frac{\chi}{m}\frac{1}{\chi} \frac{\partial \Delta F}{\partial \mu}= 
\cdots +\frac{\chi}{8m^2}\TO \sum_q \bq^2 P(q), 
$$ 
which is exactly $-\d D_{d2}$. The same holds for 
$\d D_{d1}$, $\d D_{d3}$ and the contribution arising by differentiating
$\chi\Om^2$ and $\rho_n(T)$ (at finite temperature) in Eq. \pref{freen}, 
leading to a cancellation of $\d D_\mu$ with the diagrams of Fig. \ref{chem}.  

\subsection{One-loop corrections at $T=0$} 
\label{tizero}
The cancellation of fermionic and mixed contributions as $T\ra 0$ can be
demonstrated with lengthly but straightforward calculations, so we will not
discuss it here, where instead we analyze the behavior of the remaining
bosonic contributions of Fig. \ref{bos}, and in particular the role of
dynamics in restoring Galilean invariance in the continuum model. We note in
passing that the cancellation of the fermionic corrections has been already
observed in Ref. [\onlinecite{kwon}], where however the authors did not
include both mixed terms and corrections to $\d D$ coming from the
third-order loops in $S_3(\th)$. While neglecting mixed term at $T=0$ is
not problematic, since they cancel out at $T=0$, neglecting the dynamic
third-order vertex $3b$ of Fig. \ref{3ver} leads to neglecting the bosonic
diagrams ${\cal B} 2$ and ${\cal B} 3$ of Fig. \ref{bos}, whose role at
$T=0$ is crucial.

Roughly speaking, the diagrams 
depicted in Fig. \ref{bos}, arising from the bosonic self-energy  
$\hat \Si_B$ only, correspond to 
$$
{\cal B}1 \Ra \frac{\chi}{m^2} (\vec\nabla\th)^2 \langle(\vec\nabla\th)^2\rangle,~~~~
{\cal B}2,\,{\cal B}3 \Ra \frac{\chi^2}{m^2} (\vec\nabla\th)^2 \langle 
(\vec\nabla\th)^2(\partial_\t\th)^2\rangle, 
$$
where the angular brackets indicate the average with respect to the Gaussian 
action. As a consequence, the first diagram ${\cal B} 1$, arising from a vertex 
$\sim (\vec \nabla \th)^4$ in the effective action $S_{eff}$, is the analogous 
of the term $\sum_{\a=1}^d (\partial_\a\th)^4$ in the quantum $XY$ model
\pref{eqxy}. Indeed, the term ${\cal B} 1$ generates the correction 
\be 
\d D_{{\cal B} 1} = -\frac{1}{4m^2} 
\frac{T}{d\O}\sum_{q}\bq^2 P(q)\chi(q), 
\label{bb1} 
\ee 
where $\chi(q)$ is given by Eq. \pref{chiq}. By evaluating Eq. \pref{bb1} in 
the hydrodynamic limit, in analogy with Eq. \pref{xy}, we find that 
\be
\d D_{{\cal B}1}=-\frac{\chi}{m^2} 
\frac{T}{d\O}\sum_{\bq,\Om}\frac{\bq^2}{\chi\Om^2+D\bq^2}=
-\frac{\chi}{2m^2} 
\frac{1}{d\O} \sum_\bq \frac{\bq^2}{\sqrt{\chi D} |\bq|}[b(\b c_s|\bq|)- 
b(-\b  c_s|\bq|)] \tende{T\rightarrow 0}
-\frac{g}{\sqrt{\chi D}}\frac{\chi}{2m^2}, 
\label{bb1b} 
\ee 
which is again a {\em finite} correction, as in the quantum $XY$ model,
see Eq. \pref{xy}. By comparing Eq. \pref{xy} and \pref{bb1b}, one finds
that the strength of the term $(\vec\nabla\th)^4$ in the continuum model is
controlled by $\chi/m^2$, and not $\xi_0^2 D$, as in the $XY$ model. 
Nevertheless, at $T=0$, the result \pref{bb1b} cannot be conclusive,
since it would lead to $\rho_s\neq \rho$ at zero temperature.  Indeed, by
adding the contribution of the remaining diagrams ${\cal B}2$ and ${\cal B}3$
$$
\d D_{{\cal B} 2}+\d D_{{\cal B} 3} = \frac{1}{8m^2} 
\frac{T}{d\O}\sum_{q}\bq^2\Om^2 P^2(q)\chi^2(q), 
$$
we find that the bosonic one-loop corrections to $D$ reduce to 
\begin{equation} 
\delta D=-\frac{1}{8m^2d}\frac{T}{\Omega}\sum_{q}{\mathbf q}^2 P(q) 
 \chi(q)\left[2-\Omega_m^2 P(q)\chi(q) \right]. 
\label{dd} 
\end{equation} 
In writing Eq. (\ref{dd}) we are relying on the fact that both  
the third- and fourth-order vertices needed to calculate $\delta D$ are 
expressed in terms of the fermionic bubble $\chi(q)$, Eq. \pref{chiq}. 
Indeed, both the static $\hat\Si_B^2=((\vec\nabla\th)^2/8m) \hat{\t}_3$ and 
the dynamic 
$\hat\Si_B^1=(i\partial_\t\th/3)\hat{t}_3$ part of the ``bosonic'' self-energy 
$\hat\Si_B$ are tied to the Pauli matrix $\hat{\t}_3$, leading to the same 
fermionic bubble \pref{chiq} as a coefficient of the phase-only 
action. As observed in Refs. [\onlinecite{aitchison95,depalo}], 
this is the relevant consequence of the Galilean-invariant form of 
the bosonic $\tau_3$ term appearing in (\ref{sigre}). 
Evaluating Eq. (\ref{dd}) in the hydrodynamic limit, we find 
\bea 
\d D&=&\frac{\chi}{2m^2}\frac{T}{d\O}\sum_{\bq,\Om}\frac{\bq^2} 
{\chi\Om^2+D\bq^2} 
\left[ 2-\frac{4\chi\Om^2}{\chi\Om^2+D\bq^2}\right]=
\frac{1}{2m^2}\frac{T}{d\O} \sum_{\bq,\Om}\bq^2\frac{-\Om^2+\o_\bq^2} 
{(\Om^2+\o_\bq^2)^2}=\nonumber\\ 
&=&\frac{1}{2m^2}\frac{T}{d\O} \sum_{\bq,\Om}\bq^2\frac{(i\Om+\o_\bq)^2} 
{[-(i\Om)^2+\o_\bq^2]^2}=
\frac{1}{2m^2}\frac{T}{d\O} \sum_{\bq,\Om}\bq^2\frac{1} 
{[-i\Om+\o_\bq]^2}=\frac{1}{2m^2}\frac{1}{d\O} \sum_\bq \bq^2b'(\o_\bq), 
\label{corr} 
\eea 
where $\o_{\mathbf q}=\sqrt{D/\chi}|{\mathbf q}|$ is the sound mode, and
$b'(x)=-\beta{\rm e}^{\beta x}/({\rm e}^{\beta x}-1)^2$. As a consequence
of the addition of the ${\cal B}2$ and ${\cal B}3$ contributions, the pole
at $i\Omega_m=-\o_{\mathbf q}<0$, responsible for a finite contribution at
$T=0$ in Eq.~\pref{bb1b}, is canceled in favor of a double pole at
$i\Omega_m=\o_{\mathbf q}>0$, leading to Eq. (\ref{corr}), where the
standard Bogoljubov reduction of $\rho_s/m$ in a superfluid bosonic system
is recognized \cite{popov}. Thus, for $T\ra 0$ the integral in Eq.  \pref{corr}
vanishes, i.e., by fully including the dynamic structure of the interaction
for the phase, we obtain that in a Galilean-invariant system $\rho_s=\rho$
at $T=0$. Notice that the result \pref{corr} arises from an exact
compensation between static and dynamic bosonic vertices for phase
fluctuations, both at Gaussian level, in determining the structure of the
propagator $P(q)$, and beyond the Gaussian level, in weighting the relative
contributions of static and dynamic parts.
 
\subsection{One-loop corrections in the classical limit} 
\label{clali}
According to Eq. \pref{bb1b}, a further result of the microscopic
derivation of the effective action is a reduction of the strength of the
static interaction term $(\vec\nabla \theta)^4$ with respect to the $XY$
model, see Eq.~\pref{xy}. This reduction is 
relevant in the {\em classical} regime, where
only the correction $\d D_{\cal B}1$ to $D$ coming from the {\em static}
interaction term ${\cal B}1\propto (\chi/m^2) (\vec\nabla \theta)^4$ 
survives in Eq. (\ref{dd}), leading to
$$
\delta D\approx -\frac{\chi}{m^2 D\xi_0^{2}}\frac{T}{d\xi_0^{d-2}}, 
$$
which is {\em qualitatively} similar to the result of the classical $XY$
model, Eq. (\ref{xyclass}). To make a quantitative comparison, one can
estimate the compressibility $\chi$ in the weak- or intermediate-coupling
regime as the value of the density of states at the Fermi 
level,\cite{notachi} which, in the continuum model, is
\be 
\chi=N(\e_F)=2\int \frac{d\bk}{(2\pi)^d}\d(\e_F-\e_\bk)=2{\cal A}_d m k_F^{d-2}, 
\label{nef} 
\ee 
where $k_F$ is the Fermi wave-vector. Nevertheless, also $\rho_s$,
i.e., $\rho$ (if we neglect quasiparticle effects \cite{notabos}), 
can be defined through $k_F$, at least in the
weak-to-intermediate-coupling regime (more precisely as long as $T_c < E_F$,
i.e., $U < t$)
$$ 
\rho_s=\rho=2\int_{|\bk|<k_F} \frac{d\bk}{(2\pi)^d}=2{\cal A}_d
\frac{k_F^d}{d}.
$$ 
It then follows that the ratio between the coefficient of the static
interaction term $(\vec\nabla\theta)^4$ in the continuum or in the $XY$
model is
\be 
\frac{\chi}{m^2D\xi_0^2}= \frac{\chi}{m\rho_s\xi_0^2}=\frac{d}{(\xi_0 k_F)^2}, 
\label{estim} 
\ee 
which means that, apart from a factor $d/2$, we have 
\begin{equation} 
{\rm Classical}~{\rm Regime} \Ra 
\frac{\delta D}{\delta D_{XY}} \sim \frac{1}{\left(k_F\xi_0\right)^{2}}. 
\label{ratio} 
\end{equation} 
As a consequence, in the classical limit, the one-loop correction to $D$ 
derived within the effective action is smaller than the corresponding 
first-order correction in the $XY$ model as far as $k_F^{-1}<\xi_0$, i.e., 
in the weak- and intermediate-coupling regime for the pairing interaction. 
 
Finally, let us consider the effect of the Coulomb interaction between the  
electrons. As we showed in Sec. \ref{conticha}, the presence of a 
density-density interaction in the microscopic Hamiltonian implies the  
dressing of the coefficients of the Gaussian action by the RPA  
series of the Coulomb potential $V({\mathbf q})$. As a consequence, in the 
hydrodynamic limit, the bare compressibility  $\chi$ is replaced by the 
dressed bubble 
$\chi_{LR}(\bq)\approx 1/V(\bq)$, and the sound mode $\o_{\mathbf q}$ of Eq.  
(\ref{corr}) is converted into the  
plasma mode $\omega_{\bf q}^P$ of the $d$-dimensional system (see Eq.  
\pref{sidroc} for the three-dimensional case). In the same way, when deriving  
the anharmonic terms in ${\cal S}_{eff}$, we must now include the RPA  
density fluctuations in all the vertices for $S_3$ and $S_4$. The  
one-loop corrections to $D$ are formally identical to Eqs.  
(\ref{dd})-(\ref{corr}), with $\chi\to\chi_{LR}$. Thus at $T=0$ we recover  
again the cancellations (\ref{corr}) of the bosonic diagrams, with  
$\o_{\mathbf q}\to\omega_{\mathbf q}^P$. At the same time, since  
$\chi_{LR}({\mathbf q})$ vanishes as ${\mathbf q} \to 0$, the  
classical ($\Omega_n =0$) term in Eq. (\ref{dd}), i.e., $\d D_{{\cal B}1}$  
given by Eq. \pref{bb1b}, reads 
$$
\delta D_{LR} \approx -\frac{T}{m^2dD}\frac{1}{\O}\sum_\bq \chi(\bq)= 
-\frac{T}{m^2dD\l e^2} \frac{{\cal A}_d\zeta^{2d-1}}{2d-1}, 
$$
whereas the quantum $XY$ model leads to the same result \pref{xyclass}  
of the neutral case. Thus we can roughly estimate 
$$ 
{\rm Classical} \quad {\rm Regime} \quad {\rm LR} \Ra 
\frac{\delta D_{LR}}{\delta D_{XY}} \sim \frac{\e_F}{\e_C}\left(\frac{1} 
{k_F\xi_0}\right)^{d+1}, 
$$
where $\e_F=k_F^2/2m$ is the Fermi energy and $\e_C=\lambda e^2 k_F $ is a  
characteristic Coulomb energy scale. As a consequence, while within the 
$XY$ model the Coulomb  interaction modifies only the low-temperature 
behavior of $\rho_s$, within the continuum model it affects also the
high-temperature classical regime.  

\section{One-loop corrections to $D$ within a lattice model} 
\label{corla}
To extend the previous results to a lattice model, we need to take into
account the discussion of Sec. \ref{latti}, in which we derived the 
phase-only action starting from a nearest-neighbor tight-binding model for 
electrons in a lattice. 

\begin{figure}[ht]
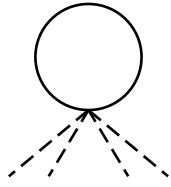
 
\begin{center} 
\insertplot{200pt}{80pt}{ }{fig10} 
\end{center} 
\caption{Fourth-order vertex of $S_4$ peculiar of the lattice case. In 
evaluating the fermionic loop one includes a  
$\partial^4\xi_\bk/\partial \bk_\a^4$ derivative of the band dispersion.} 
\label{new4ver} 
\end{figure}

At Gaussian level, the item (i) of Sec. \ref{latti} allows us to  
define the superfluid stiffness of a lattice model according to 
Eq. \pref{stiff}. In particular, a factor ${1\over 8}\Lambda_a$, with 
$$
\Lambda_a=\frac{\partial^2 \xi_{\mathbf k}}{\partial \bk_a^2}. 
$$
appears in the fermionic bubble which defines the diamagnetic term,
i.e., the coefficient of  the $(\vec\nabla \theta)^2$ term in the Gaussian
action for phase fluctuations. As a consequence, $D$ is 
controlled at $T=0$ by the mean kinetic energy 
${\cal T}$, Eq. \pref{d0}, instead of $\rho/m$ of the continuum case.  
In deriving anharmonic terms in $S_{eff}(\th)$, we must take into account 
item (i) in the proper definition of the fermionic bubbles of Figs.  
\ref{3ver} and \ref{4ver}, which give the vertices for the interacting  
phase-only model. Moreover, we need to include also the fourth-order vertex  
depicted in Fig. \ref{new4ver}, which is peculiar of the lattice case, and 
arises from the $n=4$ term of Eq. \pref{int} (see also Fig. \ref{fig3}). 
 
\begin{figure}[ht]
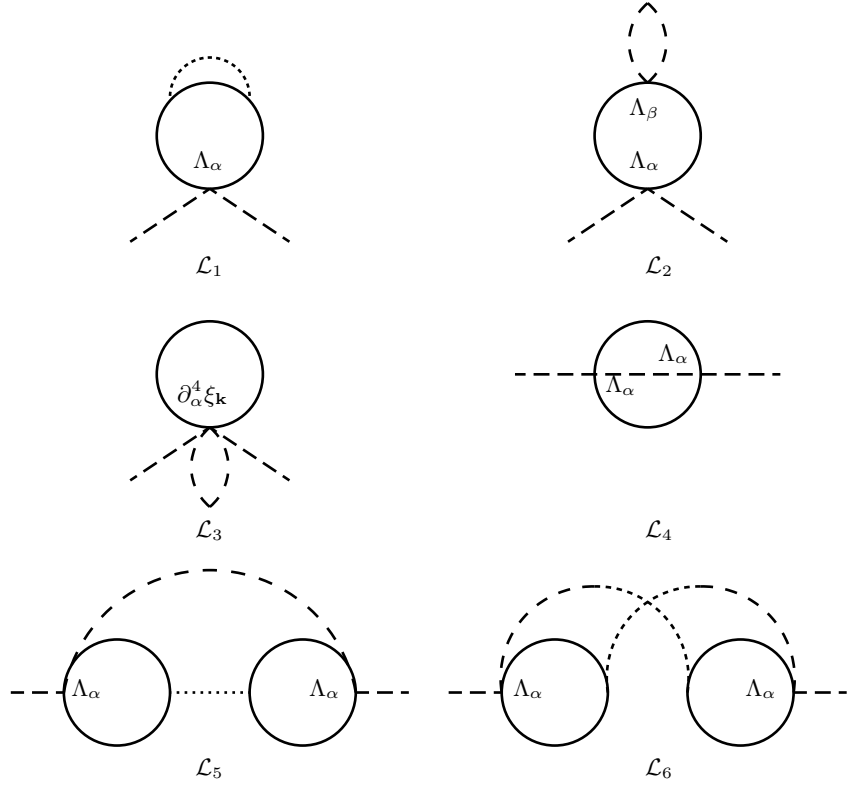
 
\begin{center} 
\insertplot{320pt}{310pt}{ 
\ins{74pt}{255pt}{$\L_\a$} 
\ins{239pt}{255pt}{$\L_\a$} 
\ins{239pt}{275pt}{$\L_\b$} 
\ins{68pt}{168pt}{$\partial_\a^4 \xi_\bk$} 
\ins{250pt}{182pt}{$\L_\a$} 
\ins{230pt}{170pt}{$\L_\a$} 
\ins{28pt}{55pt}{$\L_\a$} 
\ins{118pt}{55pt}{$\L_\a$} 
\ins{195pt}{55pt}{$\L_\a$} 
\ins{283pt}{55pt}{$\L_\a$} 
\ins{75pt}{25pt}{${\cal L}_5$} 
\ins{245pt}{25pt}{${\cal L}_6$} 
\ins{75pt}{115pt}{${\cal L}_3$} 
\ins{245pt}{115pt}{${\cal L}_4$} 
\ins{75pt}{215pt}{${\cal L}_1$} 
\ins{245pt}{215pt}{${\cal L}_2$} 
} 
{fig11} 
\end{center} 
\caption{One-loop correction to $D$ within the lattice model. As usual, we 
indicate with the dashed line the gradient of the phase and with the 
dotted line the time-derivative of the phase. Moreover, we explicitly 
indicate the insertions of band-dispersion derivatives in the fermionic 
loops.} 
\label{fig11} 
\end{figure}

At this point the calculation of one-loop correction to $D$ follows the
same steps described in Sec. \ref{corco}.  The cancellation of fermionic
and mixed one-loop corrections to $D$ at $T=0$ still holds, thus we can
again focus on bosonic corrections only.  As far as the corrections in
Fig. \ref{chem} are concerned, they are no more canceled out by the
contribution $\d D_\mu$ due to the chemical-potential shift beyond the
Gaussian level. In particular, while discussing in detail in
Sec. \ref{corco} the case of the term $d2$, we stressed that in the
continuum model such a cancellation holds because the fermionic bubble of
diagram $d2$ is the same bubble which one obtains by differentiating the
diamagnetic part of $D$ with respect to $\mu$, so that $\partial D/\partial
\mu=\chi/m$.  Indeed, since in the continuum case the factor $\L_\a=1/m$
associated to the $(\vec\nabla\th)^2$ insertion is a constant, one obtains
the same two-line fermionic loop both by expanding $S_{eff}$ up to $S_4$,
leading to $d2$, and by differentiating the single-line diagram which
defines $\rho$. However, in the lattice case the factor $\L_\a$ is not a
constant. As a consequence the diagram obtained by differentiating the
diamagnetic term ${\cal T}\equiv D(0)$ with respect to $\mu$ has {\em one}
insertion of $\L_\a$, while the the diagram $d2$, obtained by expanding
$S_{eff}$ according to item (i), has a $\L_\a$ factor for each $\th$-vertex
(diagram ${\cal L}_2$ in Fig. \ref{fig11}). The same holds for the $d1$
diagram.
 
As a consequence, in the lattice case, the contributions to $D$ come from
the bosonic diagrams depicted in Fig. \ref{fig11}, and from the
correction $\d D_\mu$ induced by the chemical-potential shift, which do not
cancel anymore the diagrams ${\cal L}_1$ and ${\cal L}_2$ in Fig. \ref{fig11}.
While evaluating $\d D_{{\cal L}2}$ we must consider that we have $2(d-1)$
bubbles with $\L_\a\neq\L_\b$ in the expansion of $S_{anh}$ in $d$
dimensions, so that
\bea 
\label{dll1} 
\d D_{{\cal L}1}&=&\frac{1}{8} \TO\sum_{q} 2P(q)\Om^2 \eta_{\rho E}(q),\\ 
\label{dll2} 
\d D_{{\cal L}2}&=&-\frac{1}{8d} \TO\sum_{q} P(q)\bq^2 [\chi_{EE}+(d-1)\tilde 
\chi_{E E}],\\ 
\label{dll3} 
\d D_{{\cal L}3}&=&-\frac{1}{8d} \TO\sum_{q} P(q)\bq^2 \tilde{\cal T},\\ 
\label{dll4} 
\d D_{{\cal L}4}&=&-\frac{1}{8d} \TO\sum_{q} 2P(q)\bq^2 \chi_{EE}(q),\\ 
\label{dll5} 
\d D_{{\cal L}5}+\d D_{{\cal L}6}&=& \frac{1}{8d} \TO\sum_{q}
P^2(q)\bq^2\Om^2 \chi_{\rho E}(q), 
\eea 
where $\eta_{\rho E}(q)$ is the fermionic bubble with three $\hat {\cal
G}_0$ lines and one insertion of $\L_\a$, and the $\chi_{ab}(q)$ bubbles
correspond to the insertion of one ($\chi_{\rho E}$) or two equal/different
($\chi_{E E}$/${\tilde \chi}_{EE}$) factors $\Lambda_\alpha$ in the
two-line bubble. As usual, when the dependence on $q$ is not explicitly
indicated, we are considering the $\bq\to 0,\Omega_m=0$ (static)
limit. Notice that, since $(\partial^4 \xi_\bk/\partial
\bk_\a^4)=-a^2(\partial^2 \xi_\bk/\partial\bk_\a^2)$, the $\tilde {\cal T}$
which appears in $\d D_{{\cal L}3}$ is proportional to the mean kinetic
energy defined by Eq. \pref{d0}, $\tilde {\cal T}=a^2{\cal T}= a^4 E_{kin}$
(the dependence on the lattice parameter $a$ is made explicit, for the sake
of clarity in the forthcoming discussion).
 
According to Eq. \pref{mush2}, the chemical-potential shift is
$$ 
\d\mu=\frac{1}{\chi}\frac{\partial \Delta F}{\partial \mu}= 
\frac{1}{8\chi}\TO\sum_q P(q)[\Om^2 \eta(q)+\bq^2\chi_{\rho E}],
$$ 
where $\eta(q)$ is the three-line bubble without factor $\L_\a$, $\chi$
is the compressibility, which carries no $\L_\a$ insertion, and
we used $\chi_{\rho E}=\partial D/\partial \mu$. By means of Eq. \pref{ddmu}, we obtain
\be 
\d D_\mu=\frac{\partial D}{\partial\mu}\d\mu=\frac{\chi_{\rho E}}{\chi} 
\frac{1}{8}\TO\sum_q P(q)[\Om^2 \eta(q)+\bq^2\chi_{\rho E}]. 
\label{ddmul} 
\ee 
Finally, the overall one-loop correction to $D$ coming from 
Eqs. \pref{dll1}-\pref{dll5} and \pref{ddmul} is 
\bea 
\delta D= 
&-&\frac{1}{8d} \frac{T}{\Omega}\sum_{q} P(q) 
{\mathbf q}^2 \left[ 2\chi_{EE}(q)+\chi_{EE}+(d-1){\tilde \chi}_{EE} 
+\tilde{\cal T}-d\frac{\chi_{\rho E}^2}{\chi}\right]- \nonumber\\ 
&- &P(q)\Omega_n^2\left[\chi_{\rho E}^2(q){\mathbf q}^2P(q)- 2d \eta_{\rho E}(q) 
+d\frac{\chi_{\rho E}}{\chi}\eta(q) 
\right]. 
\label{ddl} 
\end{eqnarray} 
 
To make a comparison with the continuum case,
Eq. \pref{corr}, we evaluate the above equation in the hydrodynamic
limit. This corresponds to calculate the $\chi_{ab}$ bubbles at zero
incoming momentum and by carefully evaluating the $\bq\ra 0$ limit for the
terms which involve $\eta_{\rho E}(q)$ and $\eta(q)$. Indeed, by simply
calculating $\sum_{\Om} P(q)\Om^2\eta(q)\ra \eta \sum_{\Om} P(q)\Om^2$ one
finds that $\sum_{\Om} P(q)\Om^2\ra\infty$. Instead, by retaining the $\Om$
dependence in the fermionic bubble one ensures the convergence of the
Matsubara sum, calculating  the $\bq\ra 0$ limit afterward.  Thus at $T=0$
limit we obtain that the overall correction to $D$ is {\em finite} in the
lattice case,
\be 
\delta D=-\frac{g}{4\sqrt{\chi D}}\left[ 3\chi_{EE}+(d-1) 
{\tilde \chi}_{EE}+\tilde{\cal T}-(d+2)\frac{\chi_{\rho E}^2}{\chi}  
\right]- \frac{g(d+1)}{\zeta\chi} 
\left[ \tilde \eta_{\rho E}-\frac{\chi_{\rho E}}{\chi}\tilde\eta 
\right], 
\label{ret} 
\ee
according to the following limiting values 
\bea 
& &\frac{T}{d\O}\sum_{\bq,\Om}\bq^2P(q)\tende{T\ra 0}\frac{2g} 
{\sqrt{\chi D}},\nonumber\\ 
& &\frac{T}{d\O}\sum_{\bq,\Om}\bq^2\Om^2P^2(q)\tende{T\ra 0}\frac{4g} 
{\chi\sqrt{\chi D}},\nonumber\\ 
& &\frac{T}{4\O}\sum_{\bq,\Om}\Om^2P(q)\eta_{\rho E}(q) 
\tende{T\ra 0}\frac{\tilde\eta_{\rho E}}{\chi}\frac{g(d+1)}{\zeta}.\nonumber 
\eea 
The $T=0$ value of $\chi$ is given by Eq. \pref{chi}. The others bubbles 
are defined as 
\bea 
\label{chiroe} 
\chi_{\rho E}&=& \frac{1}{\O}\sum_\bk  
\left(\frac{\partial^2 \xi_{\mathbf k}}{\partial \bk_\alpha^2}\right) 
\frac{\D_\bk^2}{E_\bk^3},\\ 
\label{chiee} 
\chi_{E E}&=& \frac{1}{\O}\sum_\bk  
\left(\frac{\partial^2 \xi_{\mathbf k}}{\partial \bk_\alpha^2}\right)^2 
\frac{\D_\bk^2}{E_\bk^3},\\ 
\label{chitee} 
\tilde\chi_{E E}&=& \frac{1}{\O}\sum_\bk  
\left(\frac{\partial^2 \xi_{\mathbf k}}{\partial \bk_\alpha^2}\right) 
\left(\frac{\partial^2 \xi_{\mathbf k}}{\partial \bk_\b^2}\right) 
\frac{\D_\bk^2}{E_\bk^3}, \quad (\a\neq\b), 
\eea 
and
\bea 
\label{eta} 
\tilde\eta&=& \frac{1}{\O}\sum_\bk  
\frac{\D_\bk^2\xi_\bk}{E_\bk^4},\\ 
\label{etaroe} 
\tilde\eta_{\rho E}&=& \frac{1}{\O}\sum_\bk  
\left(\frac{\partial^2 \xi_{\mathbf k}}{\partial \bk_\alpha^2}\right) 
\frac{\D_\bk^2\xi_\bk}{E_\bk^4}. 
\eea 
 
Thus in the lattice case the $T=0$ mean-field value of the  
superfluid density is corrected by one-loop phase fluctuations according to
Eq. \pref{ret}. It is interesting to observe that the continuum case is 
recovered from Eq. \pref{ret} by performing the limit $a\rightarrow 0, 
\,t\rightarrow \infty$, keeping $2a^2t=1/m$ finite. Indeed, 
since $\xi_\bk=-2t\epsilon_\bk-\mu$, with 
$\epsilon_\bk=\sum_a \cos(ak_a)$, 
$\L_\a=(\partial^2\xi_\bk/\partial k_a^2)=2ta^2 \epsilon_\bk \approx 1/m$ in 
this limit, and according to Eqs. 
\pref{chiroe}-\pref{etaroe}, one finds
$\chi_{EE},{\tilde \chi}_{EE} 
\to \chi/m^2$, $\chi_{\rho E}\to \chi/m$, 
$\tilde\eta_{\rho E}\to \tilde\eta/m$,  
while  
$\tilde{\cal T}=a^2D\rightarrow 0$.  
Thus in this limit, Eq. (\ref{ret}) correctly reduces to the continuum 
(Galilean-invariant) result  $\delta D=0$ of Eq. \pref{corr}.  
 
We now compare $\delta D$ and
$\delta D_{XY}$ in the lattice case. Differently from the continuum case,
this comparison is now meaningful both in the classical limit
and at $T=0$, since now we find finite corrections to $D$. By
comparing Eq. \pref{ret} and Eq. \pref{xy} at $T=0$, we see that
roughly speaking the strength $\sim \xi_0^2 D$ of the interaction in the
quantum $XY$ model is given by the quantity in square brackets in the first
line of Eq. \pref{ret}, i.e., it is of order 
$\sim \chi_{EE}, \tilde\chi_{EE}, \chi_{\rho E}^2/\chi$. 
Following the same arguments which led us to Eq. \pref{estim}, we can estimate
$$
\chi_{EE},\, \tilde\chi_{EE},\, 
\frac{\chi^2_{\rho E}}{\chi}\approx \frac{\xi_0^2 D}{( k_F\xi_0)^2}, 
$$
where $\chi_{EE}, \tilde\chi_{EE}, \chi_{\rho E}^2/\chi$ have the same role  
as $\chi/m^2$ in Eq. \pref{estim}, and
$k_F\sim 1/a$ in the lattice case. However, in the lattice case the dynamic 
contribution controlled by $\chi_{\rho E}/\chi^2$ does not cancel with the
static contribution, and both contribute to $\d D$ at $T=0$. 
Analogously, one finds that 
${\cal T}=a^2 D\approx D/k_F^2=\xi_0^2 D/(k_F\xi_0)^2$. Moreover, 
$$ 
\tilde \eta \simeq \frac{N(\e_F)}{1+(\mu/\D)^2}\simeq \frac{N(\e_F)} 
{( k_F\xi_0)^2} 
$$ 
where $N(\e_F)$ is evaluated in Eq. \pref{nef}, $\D$ is 
the maximum gap value, and we used the fact that at weak coupling  
$k_F \xi_0\simeq k_Fv_F/\D=\mu/\D\gg 1$.\cite{noi4} Taking into account the  
prefactor $g/\zeta\chi$ in the second line of Eq. \pref{ret}, we have
\be 
-\frac{g}{\zeta\chi} 
\left[ \tilde \eta_{\rho E}-\frac{\chi_{\rho E}}{\chi}\tilde\eta \right] 
\approx \d D_{XY} \frac{1}{(k_F \xi_0)^3} 
\label{condin} 
\ee 
where $\tilde\eta_{\rho E}\sim\tilde\eta/m$. Thus, the contributions  
coming from the three-line fermionic bubbles are subleading, and   
at $T=0$ Eq. (\ref{ret}) leads to an estimate 
\be 
\Ra \frac{\d D}{\d D_{XY}}  \sim \frac{1}{\left(k_F\xi_0\right)^{2}}. 
\label{ratioret0} 
\ee 
of the same order of Eq. (\ref{ratio}), within a numerical factor.
The presence of Coulomb forces does not change qualitatively this
conclusion, and introduces only minor quantitative corrections. This is due
to the fact that the RPA expressions for $\chi_{EE},\tilde\chi_{EE}$ have a
{\em finite} limit for ${\mathbf q}\to 0$, contrary to the continuum case,
\be
\label{chieelr}
\chi_{EE}^{LR}=\chi_{EE}-\frac{\chi_{\rho E}^2V({\mathbf q})} 
{1+V({\mathbf q})\chi} \simeq  
\left(\chi_{EE}-\frac{\chi_{\rho E}^2}{\chi}\right) 
+\frac{\chi_{\rho E}^2}{\chi}\frac{|{\mathbf q}|^{d-1}}{\lambda e^2}. 
\ee
When $\d D$ is evaluated with the long-range phase propagator $P(q)$, one
sees that the leading terms in powers of $1/(\xi_0 k_F)$ come from the
finite $\bq=0$ limit of $\chi_{EE},\tilde\chi_{EE}$ in
Eq. (\ref{chieelr}), and from the $\tilde{\cal T}$ term, 
all of order $1/(\xi_0 k_F)^2$ with respect to the coefficient 
$\xi_0^2 D$ of the $XY$ model, whereas
$$
\chi_{\rho E}^{LR}=\frac{\chi_{\rho E}}{1+V(\bq)\chi_{\rho E}} \tende{\bq
\ra 0} \frac{1}{V(\bq)},
$$
so that dynamical terms are subleading at $T=0$ in the long-range
case. This is clearly true as far as the constant in the parenthesis of
Eq. (\ref{chieelr}) is a finite quantity. Indeed, when it vanishes by
approaching the continuum limit, all (static and dynamics) terms are of the
same order, canceling each other as discussed above.

Finally, the same result (\ref{ratioret0}) holds in the classical limit,
both for the neutral and for the charged system. As we discussed above, at
high temperature only the static correction to $D$ contributes, with a
coefficient controlled in the neutral case by $\chi_{EE},
\tilde\chi_{EE},\tilde {\cal T}$, or by their $\bq=0$ finite limit in the
charged case, which play the same role (and have the same estimate) as
$\chi/m^2$ in the continuum case.
 
As far as the effect of bosonic degrees of freedom is concerned, our result
\pref{ratioret0} suggests that the $XY$ model can eventually lead to an
{\em overestimate} of phase-fluctuation effects with respect to the result
obtained by means of the microscopically derived effective action. Indeed,
according to Eq. (\ref{ratioret0}) the depletion of $D$ due to phase
fluctuations can be quite small at weak and intermediate coupling
(particularly in the BCS limit) both in the quantum and in the classical
regime.
 
On the other hand, as far as the temperature dependence of $\d D$ for
temperatures $0<T<T_{cl}$ is considered, one should take into account also
the effects of fermionic and mixed contributions. As we discussed above,
quasiparticles contribute to the depletion of $D$ at finite temperature
already at Gaussian level, with a temperature dependence determined by the
symmetry of the order parameter, which is linear in $T$ in $d$-wave
superconductors. One would expect that at $T>0$ the one-loop corrections to
$D$ coming from fermionic and mixed diagrams correct quantitatively but not
qualitatively the $D(T)$ dependence due to quasiparticles at Gaussian
level. However, a detailed analysis of the temperature dependence of $D(T)$
as arising from bosonic, mixed and fermionic contribution at low
temperature is a much harder task.  From one side, one has to face the
difficulty of evaluating the $q=0$ limit of these expressions, as we
discussed in Sec. IV A. On the other side, at finite temperature, non-local
terms appear when deriving the phase-only action, making the hydrodynamic
expansion not well-defined.  These are the so-called Landau singularities,
arising from the non analytic behavior at small $q$ of the
$\L_{JJ}$.\cite{sharapov} In any case the analysis of these issues requires
separate investigation and is not addressed in the present work.

\section{The strong-coupling regime} 
\label{stronco}
The results \pref{ratioret0} for the lattice case was 
derived by assuming $1/(k_F \xi_0)$ as a small expansion parameter.
In particular, this assumption allowed us to estimate the 
fermionic bubble according to Eq. \pref{nef}, i.e., $\chi \simeq N(\e_F)$, 
and to neglect  the contribution to $\d D$ of the  
$\tilde \eta_{ab}$ bubbles, see Eq. \pref{condin}.
One should expect that such approximations are reliable in the weak- or  
intermediate-coupling regime, where the correlation length $\xi_0$ is 
larger than the lattice parameter $a$, which sets the scale of the Fermi  
wave-vector $k_F\sim 1/a$. However, in the strong-coupling regime for the 
pairing interaction several modifications should be included. The 
fermionic bubbles can no longer be estimated as the density of states at 
the Fermi level; the correlation length is generically expected to 
be of the same order of magnitude as the 
lattice parameter; one should include both the RPA fluctuations, 
also induced by $U$, in the particle-hole channel, and the fluctuation 
of $|\Delta|\simeq(U/2)\sqrt{\rho(2-\rho)}$, which fluctuates because $\rho$  
fluctuates.\cite{depalo} This last issue can be better understood by
analogy with the superfluid bosonic systems, where the order
parameter is directly the square-root of the density, and one expects that,
in the strong-coupling regime, the fermionic system can be mapped onto an
effective bosonic system, where Cooper pairs act as bosons with a weak
residual repulsion. The above points are intimately connected, as we 
already discussed in Ref. \onlinecite{noi4}. There, the behavior of the
correlation length, properly defined as the characteristic length scale
for the fluctuations of the modulus of the superconducting order parameter,
was analyzed in the strong-coupling regime. The main finding is that
$\xi_0$ attains a value of the order of the lattice parameter $a$ at finite
densities, and diverges as $1/\sqrt{\rho}$ in the low-density regime 
$\rho\to 0$, recovering the well-known behavior of a bosonic system.\cite{bos} 

In the following we consider the $s$-wave case, which allows us for a
transparent analytical treatment. We first address the neutral case,
i.e. the model \pref{h0}-\pref{hi} with $w(\bk)=1$ (negative-$U$ Hubbard
model), and then we extend our results to the charged case. In the presence
of $d$-wave pairing the analysis becomes more difficult, even though the
physics is not expected to be different from the $s$-wave case, as we
briefly discuss at the end of this section.
 
As we said at the beginning of Sec. \ref{contine}, the 
Hubbard-Stratonovich decoupling
is a useful tool which allows us to substitute the microscopic model with
an effective action written in terms of the relevant collective degrees of
freedom. However, the choice of the relevant variables depends on the
physical properties of interest and on the energetic scales involved in the
problem. For example, the negative-$U$ Hubbard model can be decoupled both
in the particle-particle channel, as we did, and in the particle-hole
channel. The latter decoupling would lead to an effective action 
depending on the
particle density as well, 
$S_{eff}=S_{eff}(\th,\d|\D|, \d\rho)$.\cite{notaspin} 
The resulting action is formally the same
that we obtained by decoupling the additional interaction Hamiltonian
\pref{eqcoul} in the particle-hole channel, but with $V(\bq)$ replaced by
the same $-U/2$ which appears in the Hubbard term. The relevance of density
fluctuations within the Hubbard model at strong coupling is addressed in
Refs. [\onlinecite{depalo,noi4}]. In deriving $S_{eff}$ up 
to Gaussian level a term proportional to the product of amplitude and density
fluctuation, $\sim \d|\D| \d\rho$, appears. The coefficient of such a
term in the
Gaussian action is vanishingly small at weak coupling. This means that
density and modulus fluctuations decouple at weak and intermediate
coupling, and one can safely neglect modulus fluctuations without loosing
information on the contribution of the density fluctuations. As a
consequence, one can derive the phase-only action according to
the procedure described in the previous sections. However, at strong
coupling the situation is more involved: as it is shown in
Refs. [\onlinecite{depalo,noi4}], by increasing $U$ 
the density-modulus coupling become
sizable, and one finds that the density of particles and the modulus of the
order parameter experience the same fluctuations. Since
the phase mode has the same behavior of the density mode, in the
strong-coupling regime we should also take into account amplitude
fluctuations to recover a consistent description of the phase fluctuations.
 
In Sec.\ref{conticha} we found that in deriving the effective 
action for the phase of
the order parameter the inclusion of density fluctuations reflects in the
replacement of the ``bare'' coefficients $\L$ of the Gaussian action with
their counterparts ${\cal L}$ evaluated within RPA resummation in the
particle-hole channel. Beyond Gaussian level one expects the RPA dressing
of the various fermionic loops which appear as coefficients of the
effective action. As far as the $\d D$ correction in Eq. \pref{ret} is
concerned, this leads, e.g., to substitute the $\chi_{ab}$ fermionic
bubbles with the RPA resummation with the potential $-U/2$ of the
corresponding density or current irreducible correlation functions
$\chi_{ab}^{irr}(q)$. We are then left with the problem of including
modulus fluctuations. In principle, to take them into account one should
derive the Gaussian effective action $S_G(\th,\d|\D|,\d\rho)$ by including
all the fields and then integrate out modulus fluctuations as well, see,
e.g., Refs. [\onlinecite{depalo,noi4}].  However, the contribution of amplitude
fluctuations to the static long-wavelength limit of the fermionic bubbles
$\chi_{ab}^{irr}(\bq \ra 0,\o_n=0)$ can be easily derived in a different
way. Let us consider, e.g., the density-density correlation function
$\chi^{irr}$. Since $\chi^{irr}=\partial \rho/\partial \mu$ (see
also Sec. \ref{corco}), we can deduce it from the mean-field 
expression \pref{rhomf}
of the density of particles, by taking into account the dependence on $\mu$
of the gap amplitude $|\D|(0)\equiv\D_0$ itself, which appears in the
definition of $E_{\bk}=\sqrt{\xi_\bk^2+\D_0^2}$ (see also Eqs.
\pref{ff0}). At weak coupling $\partial \D_0/\partial \mu$ can be
neglected, and one finds the definition
\pref{chi} of $\chi^{irr}$. At strong coupling, where
modulus and density fluctuations are proportional, by working at fixed
number of particles the chemical-potential variations needed to preserve
the particle number reflect into amplitude variations, 
and $\partial \D_0/\partial\mu$ gives a significant contribution 
to the irreducible bubbles. We then have (at $T=0$):
$$
\chi^{irr}=\frac{\partial\rho_{MF}}{\partial\mu}=\frac{1}{\O} \sum_{\bk} 
\frac{\D_\bk^2}{E_\bk^3} + \frac{1}{\O} \sum_{\bk}\frac{\xi_\bk}{E_\bk^3} 
\D_0 \frac{\partial \D_0}{\partial \mu}, 
$$
where $\partial\D_0/\partial\mu$ can be obtained from the self-consistency 
saddle-point equation 
\be 
\frac{2}{U}=\frac{1}{\O} \sum_{\bk}\frac{1}{E_\bk}, 
\label{sceq} 
\ee 
by differentiating both sides with respect to $\mu$,
$$
\D_0 \frac{\partial \D_0}{\partial \mu}=\frac{\sum_{\bk}(\xi_\bk/E_\bk^3)} 
{\sum_{\bk}(1/{E_\bk^3})},
$$
so that
\be 
\chi^{irr}=\frac{1}{\O} \sum_{\bk} 
\frac{\D_\bk^2}{E_\bk^3} + \frac{1}{\O} \frac{\left[\sum_{\bk} 
(\xi_\bk/E_\bk^3)\right]^2}{\sum_\bk (1/E_\bk^3)}. 
\label{cirr} 
\ee 
 
In the limit of strong pairing interaction, we can evaluate
the previous bubble for $t/U\ll 1$, by taking into account the
strong-coupling solution of the coupled saddle-point equations for the
amplitude $\D_0$ Eq. \pref{sceq} and for the particle number
Eq. \pref{rhomf}, 
\begin{eqnarray*} 
\Delta_0&=&\frac{U}{2}( 1-d\alpha^2)\sqrt{1-\delta^2}+ 
{\cal O}\left({t^4\over U^3}\right),\\ 
\mu&=&-\frac{U}{2}\delta (1+2d\alpha^2)+{\cal O}\left({t^4\over U^3}\right), 
\end{eqnarray*} 
where we introduced the short-hand notation $\d\equiv 1-\rho$ and the small
parameter $\a\equiv 2t/U$. We then obtain
$$
\chi^{irr}=\frac{2}{U}\left(1-2d\a^2\right).
$$
Thus, including also the RPA resummation of $\chi^{irr}$ in the
particle-hole channel, we finally get
\be
\chi=\frac{\chi^{irr}}{1-(U/2)\chi^{irr}}=\frac{1}{2d\a t}.
\label{chirpa}
\ee 
The expression \pref{chirpa} for the density-density
electronic correlation function at $q=0$ is the same which one would obtain
by explicitly integrating out both the
$\d\rho$ and the $\d|\D|$ field  in the action 
$S(\th,\d|\D|,\d\rho)$. The latter procedure was indeed adopted in
Ref. [\onlinecite{noi4}] while deriving the propagator 
for the Hubbard-Stratonovich
field associated to the density. The result (\ref{chirpa}) shows a large
increase of $\chi$ at large $U$: this could be expected, since the
electrons are strongly paired, naturally increasing the compressibility of
the system.
 
As far as the other fermionic bubbles are concerned, we follow the 
same steps. In analogy with the previous discussion
\bea 
\chi_{\rho E}^{irr}&\equiv& \frac{1}{d}\frac{\partial D}{\partial\mu},\nonumber\\ 
\chi_{EE}^{irr}+(d-1)\tilde\chi_{EE}^{irr}&\equiv& t\frac{\partial D_{EE}}{\partial
t},\nonumber
\eea 
where the stiffness is defined in Eq. \pref{stiff}, and
$$ 
D_{EE}=\frac{1}{2t\O} \sum_\bk  
\left(\frac{\partial^2 \xi_{\mathbf k}}{\partial \bk_\alpha^2}\right) 
\left(1-\frac{\xi_\bk}{E_\bk}\right),
$$ 
at $T=0$. By taking the derivative with respect to the hopping $t$ one
includes also $\partial \Delta_0/\partial t$, as obtained again from Eq. (\ref{sceq}).
In the strong-coupling limit
\bea 
\label{rr} 
\chi_{\rho E}^{irr}&=&-4\a^2 \d,\\ 
\label{ee} 
\chi_{EE}^{irr}&=&2t\a(1-\d^2),\\ 
\label{eep} 
\tilde\chi_{EE}^{irr}&=&{\cal O}(t\a^3).
\eea 
By introducing the dressed 
potential $\tilde U=(U/2)/[1-(U/2)\chi^{irr}]=U/4d\alpha^2$, 
in analogy with Eq. (\ref{chirpa}), we find
\bea 
\label{cresc}
\chi_{\rho E}=\chi_{\rho E}^{irr}[1+\chi_{\rho\rho}^{irr}\tilde U]= 
-\frac{2\d}{d},\\
\label{ceesc} 
\chi_{EE}=\chi_{EE}^{irr}+(\chi_{\rho E}^{irr})^2\tilde U= 
2t\a(1-\d^2)+\frac{8t\a}{d}\d^2,\\ 
\label{cteesc}
\tilde\chi_{EE}=\tilde\chi_{EE}^{irr}+(\chi_{\rho E}^{irr})^2\tilde U= 
\frac{8t\a}{d}\d^2.
\eea 
The inclusion of the modulus fluctuations always 
adds a second  term to the weak-coupling definition of the 
irreducible bubbles, see, e.g., Eq. \pref{cirr}. 
This additional term is never of lower order in $\alpha$
with respect to the term obtained with the weak-coupling definitions 
(\ref{chiroe})-(\ref{chitee}). 
Since one expects that the same holds for the $\eta$ bubbles, let us first
estimate the RPA resummation of the weak-coupling definitions (\ref{eta}) and
(\ref{etaroe}), which do not include modulus-fluctuation contribution. 
The resulting corrections to $D$, arising from the second line
of Eq. (\ref{ret}), are subleading with respect to the contribution coming 
from the $\chi_{ab}$ bubbles, first line of Eq. (\ref{ret}). Since
modulus-fluctuation contributions do not change this result, we can neglect
the correction $\delta D$ coming from the $\eta$ bubbles.

By using the expressions (\ref{cresc})-(\ref{cteesc}), the
strong-coupling expression
$$
D=a^{2-d}2t\alpha (1-\delta^2),
$$
and the fact that $\tilde {\cal T}=a^2 D$, we find that Eq. (\ref{ret}),
at large $U/t$, reduces to
\begin{equation}
\delta D= -\frac{g}{\sqrt{\chi D}} 8a^{4-d} t\alpha(1-\delta^2)=
-\frac{4g}{\sqrt{\chi D}}a^2 D,
\label{dsclt}
\end{equation}
where we reintroduced explicitly the dependence on the lattice parameter
$a$, to allow one for a direct comparison with the result of the 
$XY$ model, Eq. (\ref{xy}). We then obtain that
\begin{equation}
\delta D=\delta D_{XY} \left(\frac{a}{\xi_0}\right)^2.
\label{sc}
\end{equation}
By taking into account the previous results for $\xi_0$,\cite{noi4} we
find that, since at finite densities $\xi_0\sim a$, the $XY$ 
model appears as the proper ``strong-coupling'' effective quantum model 
for phase fluctuations in a lattice. A noticeable exception is found
in the extreme low-density regime, where $\xi_0$ 
diverges as $1/\sqrt\rho$.
Moreover, since $g\sim 1/\xi_0^{d+1}$, Eq. (\ref{dsclt}) shows 
that in the low-density limit
the relative superfluid-stiffness correction $\d D/D$ vanishes as
$\rho^{d/2}$. Observe that instead the $XY$ model would lead to a
relative correction $\d D_{XY}/D \sim \rho^{d/2-1}$ which does not vanish
at $\rho=0$ in $d=2$. 

In the classical regime, as we already discussed in the weak-coupling
case, only static anharmonic terms survive, leading to
$$
\frac{\delta D}{\delta D_{XY}}=\left[1+\frac{2\delta^2}{d(1-\delta^2)}\right]
\left(\frac{a}{\xi_0}\right)^2.
$$
Since the quantity in parenthesis behaves as the microscopically derived
$(\xi_0/a)^2$ at all densities,\cite{noi4} up to a numerical constant of
order one, we find that the in the classical regime the $XY$ model turns
out to be appropriate, regardless of the density. Observe that the
numerical difference comes out from the fact that the $XY$ model attributes
a somewhat arbitrary coefficient to the anharmonic term
$(\vec\nabla\theta)^4$. If one assumed in Eq. (\ref{eqxy}) higher order
harmonics as
$$\frac{D_n}{4}\sum_{<i,j>}(1-\cos \theta_{ij})^n, \quad (n\geq 2), 
$$
which do not modify the Gaussian term, by adjusting the $D_n$ coefficients
one could reproduce the correct $(\vec \nabla \theta)^{2n}$ term as it
arises in the microscopic expansion.  In such a case the classical limit of
this extended $XY$ model and of the microscopic model would lead to exactly
the same result. As it has been noted recently in Ref. \cite{kim}, the 
mapping of the classical phase-only action of a lattice systems on an 
extended $XY$ model can also lead in two dimensions to an 
enhanced fluctiation region near the Kostelitz-Thouless transition.

Let us consider now the case when also the Coulomb potential is
present. The effect of modulus fluctuations on the irreducible fermionic
bubbles is the same as already discussed above , so Eqs. \pref{cirr} and
\pref{rr}-\pref{eep} are still valid. As far as density fluctuations are
concerned, we just have to take into account that the RPA resummation of
the irreducible bubbles in the particle-hole channel must be performed with
the full potential $V(\bq)-U/2$.  As a consequence, even though we are
considering a strong-coupling (i.e. large $U$) limit, as $\bq$ goes to zero
the Coulomb interaction $V(\bq)$ is always predominant. Thus in the
hydrodynamic limit the contributions to the correction $\d D$ only come
from the static interaction vertices, as already discussed in the
weak-coupling limit, since the $\chi_{\rho E}$ bubble vanishes for small
$\bq$, while the current-current correlation functions attain a finite
value.  More explicitly, by performing the RPA resummations
\pref{cresc}-\pref{cteesc} with $\tilde U$ replaced by the full dressed
potential $\tilde U\rightarrow \tilde
V(\bq)=-[V(\bq)+U/2]/[1+(V(\bq)-U/2)\chi^{irr}]$, it can be easily seen
that 
\bea 
\chi_{\rho E} &\tende{\bq \rightarrow 0}& \frac{1}{V(\bq)},\nonumber\\
\chi_{EE}&\tende{\bq \rightarrow 0}& \chi^{irr}_{EE}- \frac{(\chi_{\rho
E}^{irr})^2}{\chi^{irr}}=2t\a (1-\d^2)+{\cal O}(\a^3t),\nonumber\\
\tilde\chi_{EE}&\tende{\bq \rightarrow 0}& \tilde\chi^{irr}_{EE}-
\frac{(\chi_{\rho E}^{irr})^2}{\chi^{irr}}={\cal O}(\a^3t).\nonumber 
\eea

As a consequence, in Eq. \pref{ddl} for $\d D$ a coefficient
$(3\chi^{irr}_{EE}+\tilde {\cal T})/8=a^2 D/2$ survives. This coefficient
plays the same role as $\xi_0^2 D/2$ in the $XY$-model, see
Eq. \pref{xy} (which however should be evaluated with the long-range
propagator $P(q)$). Thus, Eq. \pref{sc} still holds, both at zero
temperature and in the classical regime, where by definition only
static anharmonic terms survive.  Observe that this result is again
not trivial, because only properly including modulus fluctuations one
can find the equivalence between $\chi^{irr}_{EE},
\tilde {\cal T}$ and $a^2 D$ in the strong-coupling regime for the pairing
potential $U$.

Finally, we briefly comment on the extension of the previous results
to the case of $d$-wave symmetry of the order parameter. As far as modulus
fluctuations in the particle-hole channel are concerned, in the
$d$-wave case this would reflect in a Hartree-Fock-like correction to
the band dispersion. Since this makes the analytical treatment not
viable, both for the coherence length and for $\delta D$ itself we
do not address this issue in the present work. However, at generic
filling the physics is not expected to be different from the $s$-wave
case, and the strong-coupling mapping of the effective action in the
$XY$ model should be preserved.

\section{Conclusions}
\label{concl}
In this paper we analyzed in detail the issue of the collective-mode
description of a superconductor in the quantum regime. In particular, we
addressed the issue from the point of view of the phase-fluctuation
correction to the superfluid density, and we extended the previously known
results for continuum microscopic models to the more realistic lattice
case.

Usually, the evaluation of the phase-fluctuation contribution to the
superfluid-density depletion is addressed within the quantum
generalizations of the classical $XY$ model, which however fails in
providing a complete description of dynamical effects. Specifically, in the
continuum system the finite correction to $D$ down to $T=0$ derived within
the $XY$ model implies the explicit violation of Galilean invariance. As we
showed, the inadequacy of the quantum $XY$ model can be understood as one
derives the phase-only action starting from the microscopic model. This
approach allows us to treat on the same footing both static and dynamic
interaction terms in the phase field: besides the {\it classical} (static)
anharmonic terms $(\vec\nabla \theta)^4$, one has third- and fourth-order
{\it quantum} (dynamic) interaction terms which contain the time derivative
of $\theta$. These quantum terms are absent in the quantum $XY$ model,
where the dynamics only appears at the Gaussian level, and induce a
correction to $D$ which cancels exactly, in the continuum case, the
contribution due to the classical interaction, restoring the equality
$\rho_s=\rho$ at $T=0$. We point out that this evident failure of the
quantum $XY$ in the continuum case naturally poses the question of the
quantitative validity of the quantum $XY$ model estimates even in the case
of a lattice microscopic model.

The detailed description of the main steps required to derive the
phase-only action was motivated by the fact that this formalism also
provides a more general framework for the description of different
microscopic systems. As a consequence, we were able to investigate the
phase mode both in the presence and in the absence of long-range Coulomb
interactions, and for both $s$-wave and $d$-wave symmetry of the order
parameter. As far as the lattice case is concerned, we also showed that the
phase-only action acquires a more complex structure due to the fact that
the minimal substitution couples the fermions to the electromagnetic field
at all the orders. Moreover, since the stiffness at $T=0$ is no longer
related to the particle density but rather proportional to the average
kinetic energy (for nearest-neighbor hopping), no conservation law is
violated by finding that classical and quantum interaction terms lead now
to a finite one-loop correction $\delta D$. When compared with the result
of the quantum $XY$ model, one also sees that $\delta D$ is of order $
1/(k_F\xi_0)^2$ with respect to $\delta D_{XY}$.  A similar result also
holds in the classical regime, where we found that the phase-fluctuation
correction to $D$ is smaller (in the weak-to-intermediate-coupling regime)
than within the classical $XY$ model by the same factor $\sim
1/(k_F\xi_0)^2$, in both the continuum and the lattice model. The reduction
of the phase-fluctuation effects for $ k_F\xi_0 \gg 1$ is made even more
pronounced in the continuum case by the inclusion of long-range Coulomb
forces.

We also devoted particular emphasis to the discussion of the
strong-coupling regime. As the interaction strength increases the fermionic
model evolves toward a bosonic one, where the Cooper pairs act as
individual bosons. Since in the superfluid the order parameter is connected
to density, one expects that the description of the strongly-interacting
electronic system requires the inclusion of both density and modulus
fluctuations.  We explicitly considered the negative-$U$ Hubbard ($s$-wave)
model. This model and its extensions are widely considered in the
literature, as candidates to capture the main features of the
intermediate-to-strong-coupling superconductors. This issue has attracted
renewed interest due to its possible relevance to the description of the
underdoped regime of the superconducting cuprates.

The inclusion of density fluctuations within the negative-$U$ Hubbard model
is performed by RPA resummation in the particle-hole channel of the
fermionic bubbles which appear as coefficients of the phase-only action. As
far as the modulus fluctuations are concerned, their inclusion in the
fermionic bubbles is directly derived by the definition of the
susceptibilities in terms of the physical quantities, although the formal
treatment of modulus fluctuation at Gaussian level is possible. The
discussion of the result strongly involves the knowledge of the behavior of
the coherence length $\xi_0$ at strong coupling as a function of the
fermion density, since it sets the spatial cut-off for phase
fluctuations. This issue was addressed elsewhere,\cite{noi4} and allows us
to conclude that at strong coupling, both in the presence and in the
absence of long-range Coulomb forces, the fermionic system maps onto the
quantum $XY$ model, except in the very low-density quantum regime.

Even though our attention was devoted to the correction to the superfluid
density, our results for the structure of the phase-only action also
suggest the possible outcomes of the analysis of phase-fluctuation effects
on other quantities. The connection between the coefficients of the
phase-only action and the various physical response functions (in specific
regimes) allows us to relate the correction to the Gaussian phase-only
action to the phase-fluctuation induced correction to the corresponding
physical quantity. This has been done, e.g., for the optical conductivity,
as discussed in Ref. [\onlinecite{noicond}] in connection to the physics of
high-$T_c$ superconductors. A second possibility concerns the
phase-fluctuation effect on the thermal conductivity, which is presently
under investigation \cite{thermal}.

\acknowledgments{This work comes out from a fruitful collaboration
with C. Castellani, to whom we are particularly indebted. We also thank 
S. De Palo, C. Di Castro, M. Grilli, S. Sharapov for many useful 
discussions and suggestions. L.B. acknowledges partial financial support by
the Swiss National Science Fundation under Grant No. 620-62868.00.
SC acknowledges financial support from the Italian MIUR, Cofin 2001,
prot. 20010203848, and from INFM, PA-G0-4.}

\appendix 
\section{Deduction of the phase-only action by means of Fermionic variable 
integrations} 
\label{dedu}

In this appendix we report some details on the derivation of the phase-only
action. As a first point we explicitly perform here  the Hubbard-Stratonovich 
transformation on the microscopic action of Eq. \pref{smicro}, 
introducing the auxiliary field  $\D(\bq, \t)$. In this way we get the 
decoupling of the interacting  part $S_I$ of  the action \pref{smicro} 
in the particle-particle channel, which now reads 
\bea 
S_I(c_{\sigma},c^{+}_{\sigma},\Delta,\Delta^{*})&=&
\frac{1}{U}\sum_{\bq,\Om}|\Delta(\bq,\Om)|^2-
\rTO\sum_{\bk,\bq,\on,\Om}\Delta(\bq,\Om)w(\bk) c^{+}_{\bk+\bqh\up}
(\on)c^{+}_{-\bk+\bqh\down}(\on-\Om)\nonumber\\ 
&-&\rTO\sum_{\bk,\bq,\on,\Om}[\Delta(\bq,\Om)]^{*}w(\bk)
c_{-\bk+\mathbf{\frac{q}{2}}\downarrow}(\on)c_{\mathbf{k}+\bqh \up}(\on-\Om).
\label{sin} 
\eea 
Here $\Delta(\bq,\Om)$ is the Fourier transform of the auxiliary
field, $\bq$ is the momentum and $\Om=2\pi m/\beta$ is the Matsubara 
frequency. 

A second  issue to be discussed is the effect of the gauge
transformation

\bea   
c_{\sigma}(\br,\tau)\longrightarrow c_{\sigma}'(\br,\tau) & = &   
c_{\sigma}(\br,\tau)e^{i\theta(\br,\tau)/2},\nonumber\\  
\label{fgauge2}  
c^{+}_{\sigma}(\br,\tau) \longrightarrow   
{c^{+}}'_{\sigma}(\br,\tau)& = &   
c^{+}_{\sigma}(\br,\tau)e^{-i\theta(\br,\tau)/2},  
\eea  

on the fermionic field appearing  $S_0$ and $S_I$. 
We start by rewriting Eq. \pref{sin} in 
real space 
\bea 
S_I(c_{\sigma},c^{+}_{\sigma},\Delta,\Delta^{*}) =  \int\int  
\frac{|\D(\bR,\tau)|^2}{U} d\bR d\tau 
& - & \int \int \D(\bR,\tau)w(\br)c^{+}_{\up}(\bR  
+\brh,\tau)c^{+}_{\down}(\bR-\brh,\tau) d\bR d\br d\tau \nonumber \\ 
& - & \int \int \D^{*}(\bR,\tau)w(\br)c_{\down}(\bR - 
\brh,\tau)c_{\up}(\bR+\brh,\tau) d\bR d\br d\tau, \label{sireal} 
\eea 
where $w(\br)=\sum_{\bk} w(\bk)e^{i\bk\br}$ is the Fourier transform of 
$w(\bk)$. After the transformation 
\pref{fgauge2}, we obtain, e.g., for the first term linear 
in $\D$ in Eq. \pref{sireal} 
\bea  
S_I & =& ...- \int_{\O}\int_{0}^{\beta}  
|\D(\bR,\tau)|e^{i\th(\bR,\tau)}w(\br)c^{+}_{\up}(\bR  
+\brh,\tau)c^{+}_{\down}(\bR-\brh,\tau) d\bR d\br d\tau - \cdots  \nonumber \\ 
\ra & \int \int_0^{\beta} & \left[|\D(\bR,\tau)| e^{i[\th(\bR,\tau)- 
\frac{\th(\bR-\brh,\tau)}{2}- 
\frac{\th(\bR+\brh,\tau)}{2}]} \,  w(\br){c^{+}}_{\up}(\bR  
+\brh,\tau){c^{+}}_{\down}(\bR-\brh,\tau)\right] d\bR d\br d\tau.  
\label{realgauge} 
\eea 
If the interaction is {\em local} in real space,
i.e., $w(\br)=\d(\br)$, as in the (isotropic) $s$-wave case, the phase $\th$
disappears from the exponential in Eq. \pref{realgauge}. In the $d$-wave
case instead, one can expand 
$e^{i[\th(\bR,\tau)-\frac{\th(\bR-\brh,\tau)}{2}-
\frac{\th(\bR+\brh,\tau)}{2}]}\simeq 1-i\br_\alpha\br_\beta
\vec{\nabla}_\alpha\vec{\nabla}_\beta~\th(\bR)+ \cdots$.  The residual
dependence on the phase affects the form of the phase-fluctuation
propagator by introducing terms which are irrelevant in the 
{\em hydrodynamical} limit. As a consequence, in the following we 
completely discard the dependence on the phase field in $S_I$, assuming
that after the gauge transformation \pref{fgauge2} $S_I$ has
the form \pref{sireal} with both $\D(x)$ and $\D^*(x)$ substituted by
$|\D(x)|$.
 
As far as $S_0$ in Eq. \pref{smicro} is
concerned, the effect of the gauge transformation can be better expressed by
first rewriting
$$
S_0=\int dx ~\phi^{+}(x)\left(\begin{array}{cc} 
\partial_{\tau} - \frac{\vec{\nabla}^2}{2m}-\mu & 0 \\ 
0 & \partial_{\tau} + \frac{\vec{\nabla}^2}{2m}+\mu \\ 
\end{array}\right)\phi(x), 
$$
where we introduced the standard Nambu spinor 
$\phi^{+}(\br,\tau)=\left(
c^{+}_{\up}(\br,\tau) ~~~ c_{\down}(\br,\tau)\right)$.
After the transformation \pref{fgauge2}, $S_0$  
gets modified into
\begin{equation}
\tilde{S_0} = S_0 + \int  dx ~\phi^{+}(x)\left\{ 
\left[\frac{i\partial_{\tau}\th(x)}{2}+\frac{(\vec{\nabla}\th(x))^2}{8m} 
\right]\hat{\tau}_3+ 
\left[\frac{i}{4m}\vec{\nabla}\th(x)\cdot  
\stackrel{\leftrightarrow}{\nabla}\right]\hat{\tau}_0\right\}\phi(x), 
\label{safter}
\ee
where the operator 
$\stackrel 
{\leftrightarrow}{\nabla}\equiv(\stackrel{\leftarrow}{\nabla}-\stackrel{\ra} 
{\nabla})$ 
acts on the fermionic variables. 
 
Introducing the Nambu notation also for the interacting part of the action, 
after the gauge transformation we find
\be
S  = \sum_{k' k}\sum_{ij}  
\phi^{+}_{i}(k ')A^{ij}_{k' k}\phi_{j}(k)+
 \sum_{q} \frac{|\D(q)|^2}{U}, 
\label{snambu} 
\ee 
where we put $k=(\bk,\on)$ and $q=(\bq,\Om)$, and the matrix
$A^{ij}_{k'k}$ can be deduced from Eqs. \pref{sin} and \pref{safter}.  As
we observed in Sec. \ref{contine}, the fermionic variables appear in the
action \pref{snambu} only in the Gaussian term, whose coefficient
$A^{ij}_{k'k}$ contains the collective variables $|\D|$ and $\th$. By
integrating out the fermions,\cite{negele}
$$ 
\int {\cal D}\phi {\cal D}\phi^{+} e^{-\sum_{k^{'}k}\sum_{ij}  
\phi^{+}_{i}(k^{'})A^{ij}_{k^{'}k}\phi_{j}(k)}=\mbox{Det} A_{k^{'}k}^{ij}, 
$$ 
and using the well-known identity: $\ln\mbox{Det} A_{k^{'}k}^{ij}= 
\mbox{Tr} \ln A_{k^{'}k}^{ij}$, where the trace $\mbox{Tr}$ acts in
the Nambu space and on the four-momenta $k,k'$, we finally get the
effective action for the bosonic degrees of freedom
\be 
S_{eff}(|\D|,\th)= \sum_{q} \frac{|\D(q)|^2}{U}-\mbox{Tr}\ln{A_{k^{'}k}^{ij}}. 
\label{sinte} 
\ee 
The second term of Eq. \pref{sinte} acquires a more readable form by 
separating the part of the matrix  $A_{k,k'}$  with the explicit structure of  
$\delta_{k k'}$  from the rest. This leads to
$$
\mbox{Tr}\ln{A_{k^{'}k}^{ij}} = 
\mbox{Tr}\ln{\left[\hat{{\cal G}}_0^{-1}-\hat{\Si}\right]}=
\mbox{Tr}\ln{\hat{{\cal G}}_0^{-1}}+ 
\mbox{Tr}\ln{\left[\hat{1}-\hat{{\cal G}}_{0}\hat{\Si}\right]}= 
\mbox{Tr}\ln{\hat{{\cal G}_{0}^{-1}}}-\sum_{N} S_{eff}^{N}, 
$$ 
where 
$$
S^{N}_{eff} = \frac{1}{N}\mbox{Tr}  
\overbrace{\left[\hat{{\cal G}_{0}}\hat{\Si}\times \cdots \times  
\hat{{\cal G}_{0}}\hat{\Si}\right]}^{N\,\mbox{ times}}, 
$$
and 
\bea 
\hat{\Si}& = & \left[\rTO\left(\frac{\on^{'}-\on}{2}\right)\th(k^{'}-k)-  
\frac{1}{8m}\TO\sum_{s}(\bk '-\bk-\mathbf{s})\cdot\mathbf{s}\th(k^{'}-k- 
s)\th(s)\right]\hat{\tau_3} \nonumber \\ 
& + & \left[\frac{i}{4m}\rTO \;(|\bk'|-|\bk|)\,(|\bk'|+|\bk|) \;\th(k^{'}-k)
\right]\hat{\tau}_0 - 
\left[\rTO w\left(\frac{\bk '-\bk}{2}\right)|\D|(k^{'}- 
k)\right]\hat{\tau_1}. 
\label{eqs}
\eea 
The BCS matrix ${\cal G}_0$ is 
\be 
\hat{{\cal G}}_{0}=\left( \begin{array}{cc} 
{\cal G}_{0}(\bk,\on) &   {\cal F}_{0}(\bk,\on) \\ 
{\cal F}_{0}(\bk,\on) &   -{\cal G}_{0}(-\bk,-\on) 
\end{array}   \right),   
\label{g00m} 
\ee 
where
\be
{\cal G}_{0}(\bk,\on)  =  \frac{-i\on- 
\xik}{\on^2+E_\bk^2}, \quad 
{\cal F}_{0}(\bk,\on) \label{ff0}  =  
\frac{w(\bk)\rTO|\D|(0)}{\on^2+E_\bk^2},  
\ee 
and $E_\bk=\sqrt{\xi_\bk^2+w^2(\bk)\D_0^2}$.
The value of the gap amplitude  $\D_0=\sqrt{T/\Omega} |\D|(0)$ in Eqs. 
\pref{ff0} is determined by the 
solution of the saddle-point equation for the effective action \pref{sinte}, 
$(\partial S/\partial |\D|(q=0))=0$,  and coincides with the mean-field value
 of the 
superconducting order parameter. The saddle point equation 
imposes a constraint only on the modulus of the order parameter, since a 
constant value of the Goldstone field $\th$ does not affect the 
action. As a consequence, ${\cal G}_0$ and ${\cal F}_0$ are exactly the BCS
 normal and
anomalous Green functions for the Hamiltonian 
\pref{h0}-\pref{hi}.\cite{schrieffer} By 
separating the field $|\D|(q)=|\Delta|(0)+\d |\D|(q)$ in its constant and 
fluctuating parts, the effective action \pref{sinte} can be expanded 
in terms of phase 
and modulus fluctuations with respect to the BCS state. Observe that since
one separates modulus and phase $|\D|(q)$ is the Fourier transform of
the real field $|\D|(x)$, and consequently
$[|\D|(q)]^*=|\D|(-q)$ and  $\sum_q |\D(q)|^2=
\sum_q ||\D|(q)|^2= \sum_q |\D|(q) |\D|(-q)$, where 
$\D(q)$ is the Fourier transform of the complex field $\D(x)$.

When, at the end, the gaussian expansion in the phase fluctuations 
around the saddle-point solution is performed, the Gaussian effective action 
reported in Eq. \pref{sgauss} is obtained, 
$$
S^G_{neutral}= \frac{1}{8}\sum_{q}\left[\Om^2 \L_{\rho\rho}(q)-
\bq_{a}\bq_{b}\L_{JJ}^{ab}(q)+2i\Om\bq_{a}\L_{\rho
J}^{a}(q)\right]|\th(q)|^2.   
$$ 
where the explicit expressions of  coefficient $\L$ are  the following

\bea  
\L_{\rho\rho}(q)& = & -\TO \sum_k {\rm tr}  
[\hat{\cal G}_0(k)\hat \t_3\hat{\cal G}_0(k+q)\hat \t_3]
\label{eqlrr}  \\  
\L_{\rho J}^{a}(q) & = &   
\TO \sum_k \frac{\bk^{a}}{m}{\rm tr}   
\left[\hat{\cal G}_{0}(\bk+\bqh,\on)\hat 
\t_0\hat{\cal G}_{0}(\bk-\bqh,\on-\Om)  
\hat \t_3\right]
\label{eqlrj}  \\
\L_{JJ}^{ab}& = & - \frac{\rho_{MF}}{m}\d_{ab}+\L^{ab}_{jj}(q),   
\label{eqljj}
\eea  
with 
\be
\L^{ab}_{jj}(q)=   
\TO \sum_k \frac{\bk^{a}}{m} \frac{\bk^{b}}{m} {\rm tr}   
\left[\hat{\cal G}_0(\bk+\bqh,\on)\hat \t_0\hat{\cal G}_0 (\bk-  
\bqh,\on-\Om)\hat \t_0\right]
\label{eqcurr}  
\ee
and
\be  
\rho_{MF}=\TO\sum_k {\rm tr}\left[\hat{\cal G}_0(k)\hat\t_3  
\right]=   
1-\frac{1}{\O}\sum_\bk\frac{\xik}{E_\bk}\tanh\left(   
\frac{\b E_\bk}{2}\right),   
\label{rhomf}  
\ee  
is the saddle-point fermion density.

In the presence of long-range Coulomb interaction, the Hubbard-Stratonovich 
transformation of the microscopic action \pref{smicro} reads 
\begin{eqnarray*}
&S_I&(c_{\sigma},c^{+}_{\sigma},\Delta,\Delta^{*},\rho_{HS})=   
\sum_{q}\left \{\frac{|\Delta(q)|^2}{U}+ 
\frac{|\rho_{HS}(q)|^2}{2V(\bq)}  
 - \D(q) \rTO\sum_{\bk,\on} 
w(\bk) c^{+}_{\bk+\bqh  
\up}(\on)c^{+}_{-\bk+\bqh\down}(\on-\Om) \right. \nonumber \\ 
&-&\left. \Delta^*(q)\rTO\sum_{\bk,\on}w(\bk)c_{- 
\bk+\mathbf{\frac{q}{2}}\downarrow}(\on)c_{\mathbf{k}+\bqh \up}(\on-\Om) 
+ i\rho_{HS}(q)\rTO\sum_{\bk,\on,\sigma}c^{+}_ 
{\bk-\bq,\sigma}(\on-\Om)c_{\bk,\sigma}(\on)\right\},\nonumber
\end{eqnarray*}
where the first, the third and the fourth terms were already present in Eq. 
\pref{sin}. The last term describes the interaction of the fermions with the 
$\rho_{HS}$. This term is not affected by the gauge 
transformation \pref{fgauge2}, since it is of the form $c^+c$. 
Integrating out the fermions the effective action for the collective  
variables\pref{sinte} is now
\be 
S_{eff}(|\D|,\th,\rho_{HS})= \sum_{q} \left\{\frac{|\D(q)|^2}{U}+ 
\frac{|\rho_{HS}(q)|^2}{2V(\bq)}\right\} 
-\mbox{Tr}\ln{A_{k^{'}k}^{ij}}, 
\label{sintecoul} 
\ee 
where the matrix elements $A^{ij}_{k'k}$ include now the 
density field $\rho_{HS}$, $
A_{k^{'}k} \longrightarrow A_{k^{'}k}+i\sqrt{T/\O}\rho_{HS}(k-k^{'}) 
\hat{\tau}_3$.
As a consequence, the density contributes to both the diagonal part of  
$A_{k'k}^{ij}$, whence 
\be 
\hat{{\cal G}}_{0}^{-1}\ra\hat{{\cal G}}_{0}^{-1}-i\rTO\rho_{HS}(0)\hat{\tau}_3, 
\label{mushift} 
\ee 
and to the self-energy matrix $
\hat{\Si} \rightarrow \hat{\Si}+i\rho_{HS}(x)\hat{\tau}_3.$
Again, we separate $\rho_{HS}(q)=\rho_{HS}(0)+\d\rho_{HS}(q)$.
The value of $\rho_{HS}(0)$ is determined by the solution of the
saddle-point equation for the action \pref{sintecoul}, i.e.  
$$
-\frac{i\rho_{HS}(0)}{V(0)}=\TO\sum_{k}\left[{\cal G}_{0}(k)+{\cal G}_{0}(-k)\right]=
\rho_{MF}
$$
where $\rho_{MF}$ is the particle density evaluated at mean-field level 
according to Eq. \pref{rhomf}.  
Therefore, the correction \pref{mushift} to the Green function is the 
standard Hartree shift of the chemical potential $\mu \ra 
\m+V(0)\rho_{MF}$.\cite{notamu}.

In the charged case, as anticipated in Sec. II B,  
in order to get the phase-only effective action at the
Gaussian level,  it is  necessary at first to evaluate 
the Gaussian expansion of $S_{eff}$  both in the 
phase and the density fluctuations
$$
S^{G}_{eff}(\th,\d\rho_{HS}) = \frac{1}{8}\sum_q \left(\begin{array}{cc}
\th(q) & \d\rho_{HS}(-q)
\end{array}\right)\hat{B}(q)\left(\begin{array}{c} 
\th(-q) \\ 
\d\rho_{HS}(q) 
\end{array}\right),  
$$
where
$$
\hat B(q)=
\left(\begin{array}{cc} 
\Om^2\L_{\rho\rho}(q)-\bq_{a}\bq_{b}\L_{JJ}^{ab}(q)+2i\Om\bq_{a}\L_{\rho  
J}^{a}& -2\bq_{a}\L_{\rho J}^{a}(q)+2i \Om\L_{\rho\rho}(q) \\ 
2\bq_{a}\L_{\rho J}^{a}(q)-2i \Om\L_{\rho\rho}(q) &  
4\left(\frac{1}{V(\bq)}+\L_{\rho\rho}(q)\right) 
\end{array}\right), 
$$ 
with the same definitions \pref{eqlrr}-\pref{eqljj} of the coefficients 
$\L$, and, in a second time, to perform the Gaussian integral over the 
field $\d \rho_{HS}$.

As a final result of this procedure, the following expression
\pref{sgaussc} for the effective action is obtained

$$
S^{G}_{charged}(\th) =  \frac{1}{8}\sum_q \left[\Om^2{\cal L}_{\rho\rho}(q)- 
\bq_{a}\bq_{b}{\cal L}_{JJ}^{ab}(q)+2i\bq_{a}\Om {\cal L}_{\rho J}^{a}(q)  
\right]\th(q)\th(-q), 
$$
where the coefficients $ {\cal L}$ are given by
\bea 
{\cal L}_{\rho\rho}(q) & = &  
\frac{\L_{\rho\rho}(q)}{1+V(\bq)\L_{\rho\rho}(q)}, \nonumber\\ 
{\cal L}_{\rho J}^{a}(q) & = & \frac{\L_{\rho  
J}^{a}(q)}{1+V(\bq)\L_{\rho\rho}(q)}, \nonumber\\ 
{\cal L}_{JJ}^{ab}(q)& = & \L_{JJ}^{ab}(q)-V(\bq)\frac{\L_{\rho  
J}^{a}(q)\L_{\rho J}^{b}(q)}{1+V(\bq)\L_{\rho\rho}(q)}.\nonumber  
\eea

\section{Effective action and gauge invariance}  
\label{inva}
In Sec. \ref{contine} we observed that the BCS bubbles
$\L(q)$ which appear as coefficients of the Gaussian effective action
\pref{sgauss} break, as it is well known \cite{schrieffer}, the
gauge invariance of the theory. In this appendix we derive the general
relationship between the coefficents $\L(q)$ and the corresponding
gauge-invariant electromagnetic response functions $K(q)$, which control
the physically accessible quantities (see also Ref. [\onlinecite{arun00, 
sharapov}]).

We first introduce the electromagnetic potential $A_\mu
=(\phi, {\mathbf A})$ into the Gaussian effective action \pref{sgauss} via
the minimal substitution $
(\partial \th/\partial t)\ra  
(\partial \th/\partial t)+ 2e\varphi$, $
\vec{\nabla}\th\ra  \vec{\nabla}\th -(2e/c){\mathbf A}$, which gives
$$
S_G(\th) \ra S_G(\th,A) = S_G(\th) + \frac{e^2}{2}\sum_{q}  
A_{\mu}(q)\L^{\mu\nu}(q)A_{\nu}(-q) 
+ \sum_{q} ie^2[q_{\mu}\L^{\mu\nu}(q)A_{\nu}(q)\th(-q)-  
q_{\mu}\L^{\mu\nu}(-q)A_\nu(-q)\th(q)], 
$$ 
where the four-dimensional notation for $\L^{\m\n}$ was introduced, so that
$\L_{\rho\rho}=\L^{00}$, $\L_{\rho J}^a=\L^{0a}$ and
$\L_{JJ}^{ab}=\L^{ab}$. We can now integrate out the $\th$ field to obtain
the partition function $Z[A_\mu]$ which allows one to define the 
electromagnetic response functions of the system as
\be 
K^{\mu\nu}(q)\equiv \left.\frac{\d^2 \ln
Z[A]}{\d A_{\mu}(q) \d A_{\nu}(- q)}\right|_{A_{\mu}(q)=0, A_{\nu}(-q)=0}=
\L^{\mu\nu}(q)+\frac{\L^{\mu\a}(q)q_{\a}q_{\b}\L^{\b\n}(
q)}{q_{\g}\L^{\g\d}(q)q_{\d}}.
\label{eqgaugein}  
\ee  
This equation provides the relation between the coefficients  
$\L^{\m\n}$ of Eq. \pref{sgauss} and the physical response functions  
$K^{\m\n}$. Now it can be easily seen that the difference between the $\L$ 
and the $K$ functions (i.e., the second term in the right-hand side of Eq. 
\pref{eqgaugein}) has a {\em purely longitudinal} structure. In other words 
phase fluctuations affect only {\em longitudinal} correlation functions
\cite{schrieffer}. 
As a consequence, we can  safely identify the {\em transverse} part of the 
Gaussian bubbles ($\L^T$) with the {\em transverse} part of the physical 
response functions ($K^T$).   
 
This fact turns out to be crucial in interpreting the Gaussian action 
\pref{sgauss} in the {\em hydrodynamic} limit, which is the physically 
interesting case while evaluating the low-temperature behavior of the 
collective mode. Indeed, by calculating the static limit $\Om=0, \bq  \ra 0$ 
of the coefficients \pref{eqlrr}-\pref{eqljj} we observe that:  
\begin{itemize}   
\item In the Gaussian action \pref{sgauss} the longitudinal part of  
$\L_{JJ}^{ab}$  appears, which, as we just showed, is not the physical  
longitudinal correlation function. However, in the $q\ra 0$ limit,  
the transverse $\L_{JJ}^T$ and longitudinal $\L_{JJ}^L$ part of  
$\L_{JJ}^{ab}$ do coincide, leading to the identification of the coefficient 
of $\bq^2$ with the stiffness, defined through the transverse 
physical function (see, e.g., Ref. [\onlinecite{arun00}]),
$$
\lim_{\bq \ra 0}\L_{JJ}^{L}(\bq,\Om=0) = \lim_{\bq \ra 0}   
\L_{JJ}^T(\bq,\Om=0)
=\lim_{\bq \ra 0}K_{JJ}^{T}(\bq,\Om=0)=- D(T).   
$$  
\item The coefficient $\L^a_{\rho J}(q)$ vanishes in the static limit, as 
it can be seen by considering its  definition \pref{eqlrj}.
\item From Eq. \pref{eqgaugein} it turns out that the coefficient   
$\L_{\rho\rho}$ coincides in the static limit with the physical   
density-density correlation function $K_{\rho\rho}$, whose static limit is   
by definition the (bare) compressibility $\k_0(T)$ of the 
system, $
\lim_{\bq \ra 0} \L_{\rho\rho}(\bq,\Om=0)=  
\lim_{\bq \ra 0} K_{\rho\rho}(\bq,\Om=0)=  
 \k_0(T)$.
\end{itemize}  
As a consequence of the above results, the Gaussian phase-only action in the  
hydrodynamic regime as exactly the expression presented in
Eq.~\pref{sgauss}.


\begin{thebibliography}{99}  

\def\prb{{Phys. Rev. B }} 
\def\prl{{Phys. Rev. Lett }} 
\def\al{{\em et al.}}
 
\bibitem{review} For a recent review see, e.g., J. Orenstein and
A. J. Millis, Science {\bf 288}, 468 (2000); V. M. Loktev, R. M. Quick, and
S. G. Sharapov, Phys. Rep. {\bf 349}, 1 (2001).
\bibitem{pf} 
E. Roddick and D. Stroud, Phys. Rev. Lett. {\bf 74}, 1430 (1995); 
V. J. Emery and S.A. Kivelson, Phys. Rev. Lett. {\bf 74}, 3253 (1995); 
E. W. Carlson, S. A. Kivelson, V. J. Emery, and E. Manousakis, 
Phys. Rev. Lett. {\bf 83}, 612 (1999). 
\bibitem{chacra} S. Chakravarty, G. L. Ingold, S. A. Kivelson, and
A. Luther, \prl {\bf 56}, 2303 (1986); S. Chakravarty, G. L. Ingold,
S. A. Kivelson, and G. Zimanyi, \prb {\bf 37}, 3283 (1988).
\bibitem{exp} W. N. Hardy, D. A. Bonn, D. C. Morgan, R. Liang, and
K. Zhang, Phys. Rev. Lett. {\bf 70}, 3999 (1993); C. Panagopoulos and
T. Xiang, Phys. Rev. Lett. {\bf 81}, 2336 (1998); Shih-Fu Lee,
D. C. Morgan, R. J. Ormeno, D. M. Broun, R. A. Doyle, J. R. Waldram, and
K. Kadowaki, Phys. Rev. Lett. {\bf 77}, 735 (1996).
\bibitem{arun00} A. Paramekanti, M. Randeria, T. V. Ramakrishnan, and 
S. S. Mandhal, \prb {\bf 62}, 6786 (2000).  
\bibitem{noi2} L. Benfatto, S. Caprara, C. Castellani, A. Paramekanti, 
and M. Randeria, \prb {\bf 63}, 174513 (2001). 
\bibitem {depalo} S. De Palo, C. Castellani, C. Di Castro, and
B. K. Chakraverty, Phys. Rev. B {\bf 60}, 564 (1999).
\bibitem{noi3} L. Benfatto, A. Toschi, S. Caprara and C. Castellani, \prb
{\bf 64}, 140506(R) (2001).
\bibitem{aitchison95} I. J. R. Aitchison, P. Ao, D. J. Thouless, and 
X. M. Zhu, \prb {\bf 51}, 6531 (1995).  
\bibitem{dattu98} D. Gaitonde, Int. Jour. of Mod. Phys. {\bf B 12}, 2717
(1998). 
\bibitem{aitchison00} I. J. R. Aitchison, G. Metikas, and D. J. Lee, \prb
{\bf 62}, 6638 (2000). 
\bibitem{kwon} H. J. Kwon, A. T. Dorsey, and P. J. Hirschfeld,
Phys. Rev. Lett. {\bf 86}, 3875 (2001). 
\bibitem{sharapov} S. G. Sharapov, H. Beck, and V. M. Loktev, Phys. Rev. B
{\bf 64}, 134519 (2001).  
\bibitem{sharapov2} S. G. Sharapov and H. Beck, \prb {\bf 65}, 134516
(2002).
\bibitem{noi4} L. Benfatto, A. Toschi, S. Caprara, and C. Castellani,
Phys. Rev. B, {\bf 66}, 054515 (2002).
\bibitem{kim} W. Kim, and P. J. Carbotte, Europhys. Lett. {\bf 59}, 761
(2002); cond-mat/0309261.
\bibitem{popov}  
For the analogy with the functional treatment of superfluid systems, see, e.g.,
V. N. Popov, {\em Functional Integrals in Quantum Field Theory and  
Statistical Physics}, Reidel, Dordrecht (1983). 
\bibitem{schrieffer} J. R. Schrieffer, {\em Theory of Superconductivity} 
Addison-Wesley (1964). 
\bibitem{notacompr} As we shall show at the end of Sec. \ref{conticha}, in the
presence of a short-range interactions $V_\bq=\tilde V$ in the
particle-hole channel the bare compressibility is corrected by the RPA
series of $\tilde V$, leading to the dressed compressibility $\k$ defined
below. Thus we define the actual compressibility $\k_0$ the ``bare'' one.
\bibitem{anderson} P. W. Anderson, Phys. Rev. {\bf 112}, 1900 (1958). 
\bibitem{notacut} One usually defines the cut-off $\zeta$ in $d$ dimensions
as ${\cal A}_d \zeta^d/d=1/\xi_0^d$, so that $(1/\O)\sum_\bq =1/\xi_0^d$.
In the following it will be relevant mainly the dependence of $g$ on the
powers of $1/\xi_0$.
\bibitem{abrikosov} A. A. Abrikosov, L. P. Gorkov, and I. E. Dzyaloshinski,
{\em Methods of quantum field theory in statistical physics}, Dover (1975). 
\bibitem{qp} See, e.g.,
P. A. Lee and X. G. Wen, Phys. Rev. Lett. {\bf 78}, 4111 (1997);  
\bibitem{samokhin} A different interpretation of the role of such
apparent divergences for a $d$-wave superconductor has been proposed
recently by K. V. Samokhin and B. Mitrovic, cond-mat/0308229.
\bibitem{notamat} The structure of the fermionic bubbles is
determined by the Pauli matrices associated to the $\th$ insertions. 
Since both bosonic insertions $\partial_\t\th$ and $(\vec\nabla\th)^2$ are 
tied to $\hat\t_3$, one recovers the same fermionic loop by inserting 
two $\partial_\t\th$ ($\ra\Om^2\th^2$ term) or two $(\vec\nabla\th)^2$ 
($\ra \bq^4\th^4$ term). As we shall show, the lattice case is
different, since each spatial derivative of the $\th$ field carries
also a same-order derivative of the band dispersion. As a consequence, 
the fermionic bubbles of $\chi$ and $d2$ have the same 
$\hat{\cal G}_0\hat\t_3\hat{\cal G}_0\hat\t_3$ structure, but different
$\bk$-dependent factors in the integrand.
\bibitem{notachi} At weak and intermediate coupling the value of $\chi$
does not change across the superconducting transition, so that $\chi$ has
the same value as in the normal state, where one easily finds $\chi=\int
d\xi N(\xi)f'(\xi)=N(\e_F)$. Nevertheless, one can find the same result by
explicitly evaluating Eq. \pref{chi} which defines $\chi$ in the
superconducting state.
\bibitem{notabos} We are discussing here bosonic corrections only, so that
we do not consider the superfluid-density depletion induced by
quasiparticle excitations at finite temperature.
\bibitem{bos} For the behavior of the correlation length in a bosonic
system see Ref. [\onlinecite{schrieffer}] and A. L. Fetter and J. D. Walecka,
{\em Quantum theory of many-particle systems}, Mc Graw-Hill, New York
(1971).
\bibitem{notaspin} 
In principle one should consider also the spin degrees of freedom, but 
these can be neglected in the low-energy limit since spin density 
fluctuations are completely decoupled from the rest of the system.  
\bibitem{noicond} L. Benfatto, A. Toschi, and S. Caprara, J. of  
Superconductivity {\bf 15}, 517 (2002).
\bibitem{thermal} A. Toschi, L. Benfatto, and S. Caprara, unpublished.
\bibitem{negele} J. W. Negele and H. Orland, {\em Quantum many-particle
system}, Addison-Wesley (1988).
\bibitem{notamu} 
{ As already discussed after Eq. (\ref{eqcoul}), the apparent divergence of  
the Coulomb interaction $V(0)=\infty$ is canceled out the ionic background.} 
 
\end{thebibliography}
\end{document}